\documentclass[showpacs,showkeys,preprintnumbers,floatfix,superscriptaddress]{revtex4}

\usepackage{graphicx}
\usepackage{graphics}
\usepackage{amsmath}
\usepackage{amssymb}

\begin{document}

\title{On the relativistic Thomas-Fermi treatment of compressed atoms and  compressed nuclear matter cores of stellar dimensions}

\author{M. Rotondo}
\affiliation{Department of Physics and ICRA, ``Sapienza'' University of Rome, P.le Aldo Moro 5, I--00185 Rome, Italy}
\affiliation{ICRANet, P.zza della Repubblica 10, I--65122 Pescara, Italy}

\author{Jorge A. Rueda}
\affiliation{Department of Physics and ICRA, ``Sapienza'' University of Rome, P.le Aldo Moro 5, I--00185 Rome, Italy}
\affiliation{ICRANet, P.zza della Repubblica 10, I--65122 Pescara, Italy}

\author{Remo Ruffini}
\email{ruffini@icra.it}
\affiliation{Department of Physics and ICRA, ``Sapienza'' University of Rome, P.le Aldo Moro 5, I--00185 Rome, Italy}
\affiliation{ICRANet, P.zza della Repubblica 10, I--65122 Pescara, Italy}
\affiliation{ICRANet, University of Nice-Sophia Antipolis, Grand Chateau, BP 2135, 28 Avenue de Valrose, 06103 Nice Cedex 2, France}
\author{S.-S. Xue}
\affiliation{Department of Physics and ICRA, ``Sapienza'' University of Rome, P.le Aldo Moro 5, I--00185 Rome, Italy}
\affiliation{ICRANet, P.zza della Repubblica 10, I--65122 Pescara, Italy}



\date{\today}

\begin{abstract}

The Feynman, Metropolis and Teller treatment of compressed atoms is extended to the relativistic regimes. Each atomic configuration is confined by a Wigner-Seitz cell and is characterized by a positive electron Fermi energy. The non-relativistic treatment assumes a point-like nucleus and infinite values of the electron Fermi energy can be attained. In the relativistic treatment there exists a limiting configuration, reached when the Wigner-Seitz cell radius equals the radius of the nucleus, with a maximum value of the electron Fermi energy $(E_e^F)_{max}$, here expressed analytically in the ultra-relativistic approximation. The corrections given by the relativistic Thomas-Fermi-Dirac exchange term are also evaluated and shown to be generally small and negligible in the relativistic high density regime.  The  dependence of the relativistic electron Fermi energies by compression for selected nuclei are  compared and contrasted to the non-relativistic ones and to the ones obtained in the uniform approximation. The relativistic Feynman, Metropolis, Teller approach here presented overcomes some difficulties   in the Salpeter approximation generally adopted  for compressed matter in physics and astrophysics. The treatment is then extrapolated to compressed nuclear matter cores of stellar dimensions with $A\simeq (m_{\rm Planck}/m_n)^3 \sim 10^{57}$ or $M_{core}\sim M_{\odot}$. A new family of equilibrium configurations exists for selected values of the electron Fermi energy  varying in the range $0 < E_e^F \leq (E_e^F)_{max}$. Such configurations fulfill global but not local charge neutrality. They have electric fields on the core surface, increasing for decreasing values of the electron Fermi energy reaching values much larger than the critical value $E_c = m_e^2c^3/(e\hbar)$, for $E_e^F=0$.  We  compare and contrast our results with the ones of Thomas-Fermi model in strange stars. 
\end{abstract}



\maketitle

\section{Introduction}\label{sec:1}

In a classic article Baym, Bethe and Pethick \cite{baym} presented the problem of matching, in a neutron star,  a liquid  core, composed of $N_n$ neutrons, $N_p$ protons and $N_e$ electrons, to the crust taking into account the electrodynamical and surface tension effects. After discussing the different aspects of the problem  they concluded: {\it The details of this picture requires further elaboration; this is a situation for which the Thomas-Fermi method is useful}. 
This statement, in first instance, may appear surprising: the Thomas-Fermi model has been  extensively applied in atomic physics (see e.g Gomb\' as \cite{gombas}, March \cite{march}, Lundqvist and March \cite{lundqvist}), also has been applied extensively in atomic physics in its relativistic form (see e.g  Ferreirinho, Ruffini and Stella \cite{ruffini}, Ruffini and Stella \cite{ruffinistella81}) as well as in the study of atoms with  heavy nuclei  in the classic works of Migdal, Popov and Voskresenskii  \cite{migdal,migdal77}. Similarly there have been considerations of relativistic Thomas-Fermi model for quark stars  pointing out the existence of critical electric fields on their surfaces (see e.g. Alcock, Farhi, Olinto \cite{alcock},  Usov \cite{usov}). Similar results have also been obtained by Alford et al. \cite{alford} in the transition at very high densities, from the normal nuclear matter phase in the core   to the color-flavor-locked phase  of quark matter in the inner core of hybrid stars. No example  exists to the application of the electromagnetic Thomas-Fermi model for neutron stars.
This problem can indeed  be approached with merit by studying the simplified but rigorous concept of a nuclear matter core of stellar dimensions which  fulfills the relativistic Thomas-Fermi equation as discussed in  \cite{rrx200613, prl}. As we will see this work leads to the prediction of the existence of a critical electric field at the interface between the core and the crust of a neutron star.

In \cite{rrx200613, prl} we have first generalized the treatment of heavy nuclei by enforcing self-consistently the condition of beta equilibrium in the relativistic Thomas-Fermi equation. Using then the existence of scaling laws we have extended the results from heavy nuclei to the case of nuclear matter cores of stellar dimensions. In both these treatments we had there assumed the Fermi energy of the electrons $E_e^F=0$. The aim of this article is to proceed with this dual approach and to consider first the case of compressed atoms and then, using the existence of scaling laws, the compressed nuclear matter cores of stellar dimensions with a positive value of their electron Fermi energies.

It is well known that Salpeter has been among the first to study  the behavior of matter under extremely high pressures by considering a Wigner-Seitz cell of radius $R_{WS}$ \cite{salpeter}.
Salpeter assumed  as a starting point the nucleus point-like and a uniform distribution of electrons within a Wigner-Seitz cell. He then considered corrections to the above model due to the inhomogeneity of electron distribution. The first correction corresponds to the inclusion of the lattice energy $E_C=-(9N_p^2 \alpha)/(10 R_{WS})$, which results from the point-like nucleus-electron interaction and, from the electron-electron interaction inside the cell of radius $R_{WS}$. The second correction is given by a series-expansion of the electron Fermi energy about the average electron density $n_e$ given by the uniform approximation. The electron density is then assumed equals to $n_e [1+\epsilon(r)]$ with $\epsilon(r)$ considered as infinitesimal. The Coulomb potential energy is assumed to be the one of the point-like nucleus with the uniform distribution of electrons of density $n_e$  thus the correction given by $\epsilon(r)$ is neglected on the Coulomb potential. The electron distribution is then calculated at first-order by expanding the relativistic electron kinetic energy about its value given by the uniform approximation considering as infinitesimal the ratio $eV/E^F_e$ between the Coulomb potential energy $eV$ and the electron Fermi energy $E^F_e = \sqrt{[c P^F_e(r)]^2 + m^2_e c^4}-m_e c^2 - e V$.
The inclusion of each additional Coulomb correction results in a decreasing of the pressure of the cell $P_{S}$ by comparison to the uniform one (see \cite{inpreparation}  for details).
 
It is quite difficult to assess the self-consistency of all the recalled different approximations adopted by Salpeter.    
In order to validate  and also to see the possible limits of the Salpeter approach,  we consider the relativistic generalization of the Feynman, Metropolis, Teller treatment \cite{fmt} which takes automatically and globally into account all electromagnetic and special relativistic contributions.    
We show explicitly how this new treatment leads in the case of atoms to electron distributions markedly different from the ones often adopted in the literature of constant electron density distributions. At the same time it allows to overcome some of the difficulties in current treatments.

Similarly the point-like description of the nucleus often adopted in literature is confirmed to be unacceptable in the framework of a relativistic treatment.

In Sec.~\ref{sec:2} we first recall the non-relativistic treatment of the compressed atom by Feynman, Metropolis and Teller. In Sec.~\ref{sec:3} we generalize that treatment to the relativistic regime by integrating the relativistic Thomas-Fermi equation, imposing also the condition of  beta equilibrium. In Sec.~\ref{sec:4} we first compare the new treatment with the one corresponding to a uniform electron distribution often used in the literature and to the Salpeter treatment. We also compare and contrast the results of the relativistic and the non-relativistic treatment.

We then proceed to analyze the case of compressed nuclear matter cores of stellar dimensions. 

In Sec. \ref{sec:5}, using the same scaling laws adopted in  \cite{rrx200613, prl} we turn to the case of nuclear matter cores of stellar dimensions with mass numbers $A\approx(m_{\rm Planck}/m_n)^3 \sim 10^{57}$ or $M_{core}\sim M_{\odot}$  where  $m_n$  is the neutron  mass and $m_{\rm Planck}=(\hbar c/G)^{1/2}$ is the Planck mass. Such a configuration present global but not local charge neutrality. Analytic solutions for the ultra-relativistic limit are obtained. In particular we find:

1) \quad explicit analytic expressions for the electrostatic field and the Coulomb potential energy, 

2) \quad an entire range of possible Fermi energy for the electrons between zero and a maximum value $(E_e^F)_{max}$, reached when $R_{WS}=R_c$, which can be expressed analytically,

3) \quad the explicit analytic expression of the ratio between the  proton number $N_p$ and the mass number $A$  when $R_{WS}=R_c$.

We turn then in Sec. \ref{sec:6} to the study of the compressional energy of the nuclear matter  cores of stellar dimensions for selected values of the electron Fermi energy. We show that the solution with $E_e^F=0$  presents the largest value of the electrodynamical structure.

We finally summarize the conclusions in Sec. \ref{sec:7}.
 
\section{The Thomas-Fermi model for compressed atoms: the Feynman-Metropolis-Teller treatment}\label{sec:2}

\subsection{The classical Thomas-Fermi model}

The Thomas-Fermi model assumes that the electrons of an atom constitute a fully degenerate gas of fermions confined in a spherical region by the Coulomb potential of a point-like nucleus of charge $+eN_p$ \cite{thomas, fermi1}. Feynman, Metropolis and Teller have shown that this model can be used to derive the equation of state of matter at high pressures by considering a Thomas-Fermi model confined in a Wigner-Seitz cell of radius $R_{WS}$ \cite{fmt}. 

We recall that the condition of equilibrium of the electrons in an atom, in the non-relativistic limit, is expressed by 
\begin{eqnarray}
\frac{(P_e^F)^2}{2m_e}-eV=E_e^F,
\label{tf1}
\end{eqnarray}
where $m_e$ is the electron mass, $V$ is the electrostatic potential and $E_e^F$ is their  Fermi energy. 

The electrostatic potential fulfills, for $r>0$,  the Poisson equation
\begin{eqnarray}
\nabla^2 V =4\pi e n_e ,
\label{tf2}
\end{eqnarray}   
where the electron number density $n_e$ is related to the Fermi momentum $P_e^F$ by
\begin{eqnarray}
n_e=\frac{(P_e^F)^3}{3 \pi^2 \hbar^3}.
\label{tf3}
\end{eqnarray} 
For neutral atoms and ions $n_e$ vanishes at the boundary so the electron Fermi energy is, respectively, zero or negative. In the case of compressed atoms $n_e$ does not vanish at the boundary while the Coulomb potential energy $eV$ is zero. Consequently $E_e^F$ is positive.  

Assuming
\begin{eqnarray}
eV(r)+E_e^F= e^2N_p \frac{\phi(r)}{r},
\label{ecn5}
\end{eqnarray}
we obtain the following expression for the electron number density
\begin{eqnarray}
n_e(\eta)=\frac{N_p}{4 \pi b^3}\left(\frac{\phi(\eta)}{\eta}\right)^{3/2},
\label{ecn7}
\end{eqnarray}
where the new dimensionless radial coordinate $\eta$ is given by $r=b\eta$, where 
\begin{eqnarray}
b=(3\pi)^{2/3}\frac{\hbar^2}{m_ee^2}\frac{1}{2^{7/3}}\frac{1}{N_p^{1/3}}.
\label{ecn8}
\end{eqnarray}
Eq. (\ref{tf2}) can be then written in the form
\begin{eqnarray}
\frac{d^2\phi(\eta)}{d\eta^2}=\frac{\phi(\eta)^{3/2}}{\eta^{1/2}},
\label{ecn11}
\end{eqnarray}
which is the classic Thomas-Fermi equation \cite{fermi1}. A first boundary condition for this equation follows from the point-like structure of the nucleus
\begin{equation}
\phi(0)=1.
\label{ecn12bis}
\end{equation}
A second boundary condition comes from  the conservation of  the number of electrons  $N_e=\int_0^{R_{WS}}{4 \pi n_e(r) r^2 dr}$
\begin{eqnarray}
1-\frac{N_e}{N_p}=\phi(\eta_0)-\eta_0 \phi'(\eta_0),
\label{ecn14}
\end{eqnarray} 
where $\eta_0=R_{WS}/b$ defines the radius $R_{WS}$ of the Wigner-Seitz cell.
\begin{figure}[th] 
\begin{center}
\includegraphics[scale=0.8]{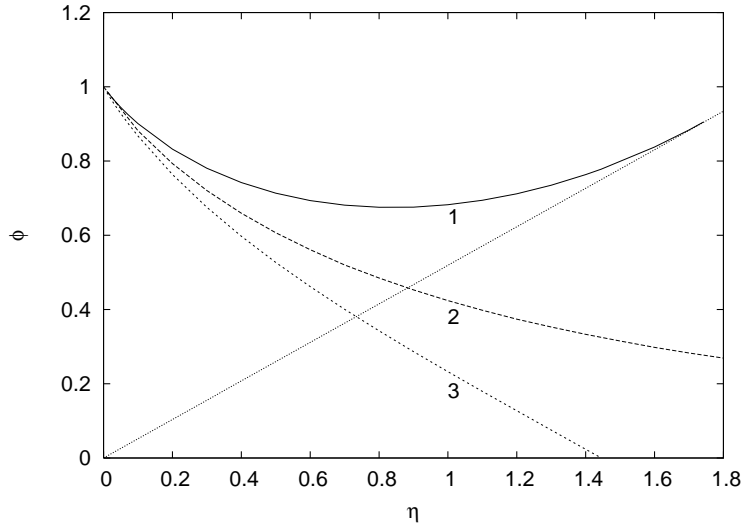}
\end{center}
\caption{Physically relevant solutions  of the  Thomas-Fermi Equation (\ref{ecn11}) with the boundary conditions (\ref {ecn12bis}) and (\ref {ecn14}).   The curve $1$ refers to a neutral compressed atom. The curve $2$ refers to a neutral free atom. The curve $3$ refers to a positive ion. The dotted straight line is the tangent to the curve $1$ at the point $(\eta_0, \phi(\eta_0))$ corresponding to overall charge neutrality (see Eq. (\ref{ecn14})).}%
\label{cap1b}%
\end{figure}
In the case of compressed atoms $N_e=N_p$ so the Coulomb potential energy $eV$ vanishes at the boundary $R_{WS}$. As a result, using Eqs. (\ref{tf1}) and (\ref{tf3}), the Fermi energy of electrons is given by
\begin{eqnarray}
E_e^F=\frac{N_p e^2}{b}\frac{\phi(\eta_0)}{\eta_0}.
\label{tf7a}
\end{eqnarray} 
Therefore in the classic treatment $\eta_0$ can approach zero and consequently the range of the possible values of the Fermi energy extends from zero to infinity.

\begin{figure}[th] 
\begin{center}
\includegraphics[scale=0.8]{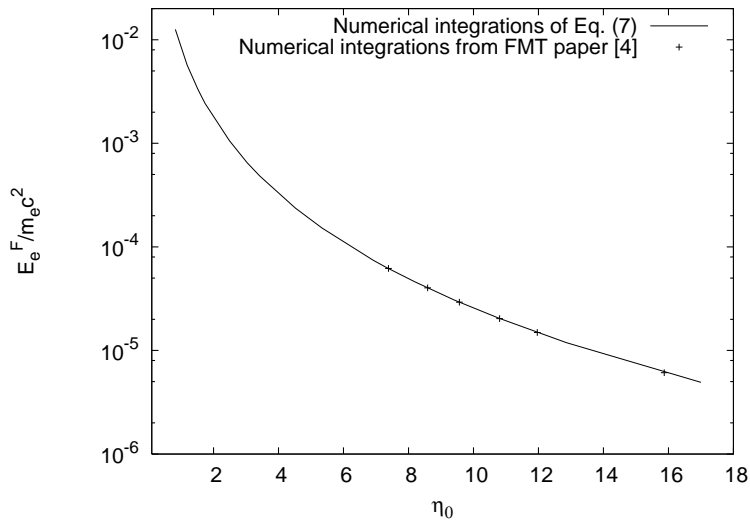}
\end{center}
\caption{The electron Fermi energies for iron, in units of the electron mass, are plotted as a function of the dimensionless compression parameter $\eta_0$. Points refer to the numerical integrations of the Thomas-Fermi equation (\ref{ecn11}) performed originally by Feynman, Metropolis and Teller in \cite{fmt}.}%
\label{cap1bc}%
\end{figure}

The results are summarized in Figs. \ref{cap1b} and \ref{cap1bc}.

\subsection{The Thomas-Fermi-Dirac model}

Dirac  has introduced modifications to the original Thomas-Fermi theory to include effects of exchange \cite{dirac30}. In this case the condition of equilibrium of the electrons in the atom is generalized as follows
\begin{eqnarray}
\frac{(P_e^F)^2}{2m_e}-eV-\frac{\alpha}{\pi} c P_e^F = E_e^F, 
\label{tf1ex}
\end{eqnarray}
where as usual $\alpha=e^2/\hbar c$ denotes the fine structure constant.

The electron number density is now connected to the Coulomb potential energy by 

\begin{eqnarray}
n_e=\frac{1}{3\pi^2\hbar^3 c^3} \left[\frac{\alpha}{\pi} m_e c^2 + \sqrt{\left(\frac{\alpha}{\pi}m_e c^2 \right)^2 + 2m_e c^2 (eV+E_e^F)}\right]^3.
\label{tf2ex}
\end{eqnarray}

Assuming

\begin{eqnarray}
\frac{1}{2}\left(\frac{\alpha}{\pi}\right)^2 m_e c^2 + eV(r)+E_e^F=e^2N_p\frac{\phi(r)}{r},
\label{tf2qex}
\end{eqnarray}

and $r=b\eta$, the Poisson equation  can be written as

\begin{eqnarray}
\frac{d^2\phi(\eta)}{d\eta^2}=\eta\left[d+\left(\frac{\phi(\eta)}{\eta}\right)^{1/2}\right]^3,
\label{ecn11ex}
\end{eqnarray}

where $b$ is given by Eq.(\ref{ecn8}) and $d=(3/(32\pi^2))^{1/3}(1/N_p)^{2/3}$. The boundary condition for Eq.~(\ref{ecn11ex}) are $\phi(0)=1$ and $\eta_0 \phi'(\eta_0)=\phi(\eta_0)$.

\section{The relativistic generalization of the Feynman-Metropolis-Teller treatment}\label{sec:3}

\subsection{The relativistic Thomas-Fermi model for atoms}

In the relativistic generalization of the Thomas-Fermi equation the point-like approximation of the nucleus must be abandoned 
\cite{ruffini, ruffinistella81} since  the relativistic  equilibrium condition 
\begin{eqnarray}
E_e^F =  \sqrt{(P_e^Fc)^2+m_e^2c^4}-m_ec^2 -e V(r)\, ,
\label{efex}
\end{eqnarray} 
which generalizes the Eq. (\ref{tf1}),  would lead to a non-integrable expression for the electron density near the origin.
Consequently we adopt an extended nucleus.
Traditionally the radius of an extended nucleus is given by the phenomenological relation $R_c=r_0A^{1/3}$ where $A$ is the number of nucleons and $r_0=1.2 \times10^{-13}cm$. Further it is possible to show from the extremization of the semi-empirical Weizsacker mass-formula that the relation between $A$ and $N_p$ is given by
\begin{eqnarray}
N_p=\left[\frac{2}{A}+\frac{3}{200}\frac{1}{A^{1/3}}\right]^{-1} ,
\label{Np}
\end{eqnarray} 
which in the limit of small $A$ gives 
\begin{eqnarray}
N_p\approx \frac{A}{2} .
\label{Np1}
\end{eqnarray} 
In \cite{prl} we have relaxed, for $E_e^F=0$, the condition $N_p\approx A/2$ (adopted, for example, in Migdal, Popov and Voskresenski \cite{migdal77}) as well as the condition $N_p=[2/A+3/(200A^{1/3})]^{-1}$ (adopted for example in Ferreirinho, Ruffini and Stella \cite{ruffini, ruffinistella81}) imposing explicitly the  beta decay equilibrium between neutron, protons and electrons.

In particular, following the previous treatments (see e.g. \cite{prl}), we have assumed a constant distribution of protons confined in a radius $R_c$ defined by
\begin{eqnarray}
R_c=\Delta \frac{\hbar}{m_\pi c}N_p ^{1/3},  
\label{pn}
\end{eqnarray}
where $m_{\pi}$ is the pion mass and $\Delta$ is a parameter such that $\Delta \approx 1$ ($\Delta < 1$) corresponds to nuclear (supranuclear) densities  when applied to ordinary nuclei. Consequently, the proton density can be written as
\begin{eqnarray}
n_p(r)= \frac{N_p}{\frac{4}{3}\pi R_c^3}\theta(R_c-r)=\frac{3}{4 \pi} \frac{m_{\pi}^3c^3}{\hbar^3}\frac{1}{\Delta^3}\theta(R_c-r),
\label{pnx}
\end{eqnarray}
where $\theta(x)$ is the Heaviside function which by definition is given by
\begin{equation}
\theta(x) = \left\{\begin{array}{ll} 0, &  \quad x<0, \\
1, & \quad x>0.
\end{array}\right.
\label{popovs1wq}
\end{equation} 

The electron density is given by
\begin{eqnarray}
n_e(r) = \frac{(P_e^{F})^3}{3\pi^2\hbar^3}=\frac {1}{3\pi^2\hbar^3c^3}\left[e^2 V^2(r)+ 2m_ec^2e V(r)\right]^{3/2},
\label{elndA}
\end{eqnarray}
where $V$ is the Coulomb potential. 

The overall Coulomb potential satisfies the Poisson equation 
\begin{eqnarray}\label{eq:eposs}
\nabla^2 V(r)= -4\pi e\left[n_p(r)-n_e(r)\right],
\label{eposs}
\end{eqnarray}
with the boundary conditions   $V(\infty)=0$ (due to global charge neutrality) and finiteness of $V(0)$. 

\begin{figure}[th]
\begin{center}
	\includegraphics[scale=0.8]{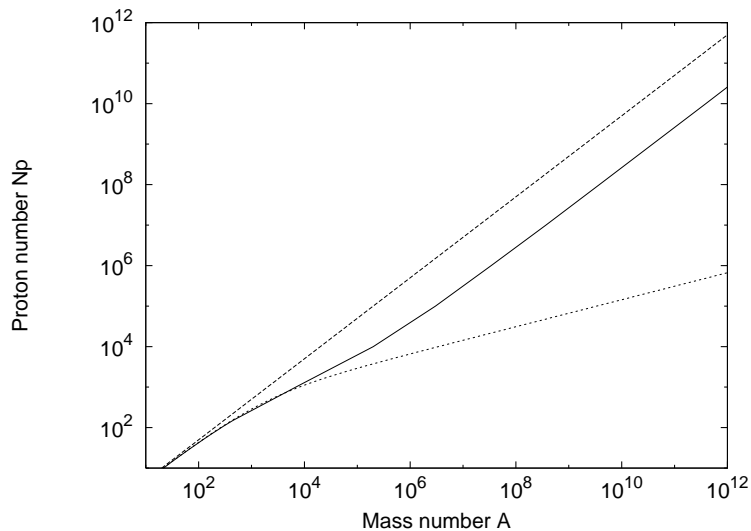}
\end{center}
\caption{ The $A$-$N_p$ relation at nuclear density (solid line) obtained from first principles compared with the  phenomenological expressions given by $N_p\simeq A/2$  (dashed line) and Eq.~(\ref{Np}) (dotted line). The asymptotic value, for $A\rightarrow (m_{\rm Planck}/m_n)^3$, is $N_p\approx 0.0046 A$.  
}%
\label{ANpx}
\end{figure}

By introducing the dimensionless quantities $ x= r/\lambda_{\pi}$, $ x_c= R_c/\lambda_{\pi}$  and $\chi/r=e V(r)/(c\hbar)$ with $\lambda_{\pi}=\hbar/(m_{\pi}c)$, and replacing the particle densities (\ref{pnx}) and (\ref{elnd}) into the Poisson equation (\ref{eq:eposs}) we obtain the relativistic Thomas-Fermi equation
\begin{eqnarray}
\frac{1}{3x}  \frac {d^2\chi(x)}{d x^2}
= -\frac{\alpha}{\Delta^3}\theta( x_c- x)
+ \frac {4\alpha}{9\pi}\left[\frac {\chi^2(x)}{ x^2}
+2\frac{m_e}{m_\pi}\frac{\chi}{x}\right]^{3/2}, \nonumber\\  
\label{eqlessx}
\end{eqnarray}
where $\chi(0)=0$, $\chi(x_{\infty})=0$.   
The neutron density $n_n(r)$, related to the neutron Fermi momentum $P_n^F=(3\pi^2 \hbar^3 n_n)^{1/3}$, is determined, as in the previous case \cite{prl}, by imposing the condition of beta equilibrium
\begin{eqnarray}
E_n^F&=&\sqrt{(P_n^Fc)^2+m^2_nc^4}-m_nc^2 \nonumber\\
&=& \sqrt{(P_p^Fc)^2+m^2_pc^4}-m_pc^2 + eV(r), 
\label{npeq1ex}
\end{eqnarray} 
which in turn is related to the proton  density $n_p$ and the electron density by Eqs.~(\ref{elndA}),  (\ref{eposs}).
Integrating numerically these equations we have obtained a new generalized relation between $A$ and $N_p$ for any value of $A$. In the limit of small $A$ this result agrees  with the phenomenological relations given by Eqs. (\ref{Np}, \ref{Np1}), as is clearly shown in Fig. (\ref{ANpx})

\subsection{The relativistic Thomas-Fermi model for compressed atoms}

We turn now to the case of compressed atoms in which the electron Fermi energy is positive.
The relativistic generalization of the equilibrium condition (\ref{tf1}) now reads
\begin{eqnarray}
E_e^F =  \sqrt{(P_e^Fc)^2+m_e^2c^4}-m_ec^2 -e V(r)> 0\, ,
\label{efe}
\end{eqnarray} 
Adopting an extended-nucleus with a radius given by Eq. (\ref{pn}) and a proton density given by Eq. (\ref{pnx}) the Poisson equation (\ref{eposs}), with the following electron density
\begin{eqnarray}
n_e(r) = \frac{(P_e^{F})^3}{3\pi^2\hbar^3}=\frac {1}{3\pi^2\hbar^3c^3}\left[e^2\hat V^2(r)+ 2m_ec^2e\hat V(r)\right]^{3/2},
\label{elnd}
\end{eqnarray}
can be written as 
\begin{eqnarray}
\frac{1}{3x}  \frac {d^2\chi(x)}{d x^2}
= -\frac{\alpha}{\Delta^3}\theta( x_c- x)
+ \frac {4\alpha}{9\pi}\left[\frac {\chi^2(x)}{ x^2}
+2\frac{m_e}{m_\pi}\frac{\chi}{x}\right]^{3/2}, \nonumber\\  
\label{eqless}
\end{eqnarray}
where $ x= r/\lambda_{\pi}$, $ x_c= R_c/\lambda_{\pi}$, $\chi/r=e\hat V(r)/(c\hbar)$, $\lambda_{\pi}=\hbar/(m_{\pi}c)$ and $e \hat V = eV+E_e^F$.  The equation (\ref{eqless}) has to be integrated with the boundary conditions $\chi(0)=0$, $\chi(x_{WS})=x_{WS}\chi'(x_{WS})$, $x_{WS}=R_{WS}/\lambda_{\pi}$. 
 
The neutron density $n_n(r)$, related to the neutron Fermi momentum $P_n^F=(3\pi^2 \hbar^3 n_n)^{1/3}$, is determined by imposing the condition of beta equilibrium
\begin{eqnarray}
E_n^F &=& \sqrt{(P_n^Fc)^2+m^2_nc^4}-m_nc^2 \nonumber\\
&=& \sqrt{(P_p^Fc)^2+m^2_pc^4}-m_pc^2 + eV(r)+E_e^F.
\label{npeq1e}
\end{eqnarray} 

Using this approach, it is then possible to determine the beta equilibrium nuclide as a function of the density of the system. Infact, as suggested by Hund \cite{hund} and Landau \cite{landau38}, when the electron Fermi energy is sufficiently high, electrons can be absorbed by protons and converted to neutrons in inverse beta decay $p+e^-\rightarrow n+\nu_e$ because the condition $E_n^F<\sqrt{(P_p^Fc)^2+m^2_pc^4}-m_pc^2 + eV(r)+E_e^F$ holds. The condition of equilibrium (\ref{npeq1e}) is  crucial, for example, in the construction of a self-consistent equation of state of high energy density matter present in white dwarfs and neutron star crusts \cite{inpreparation}. In the case of  zero electron Fermi energy the generalized $A-N_p$ relation of Fig. (\ref{ANpx}) is obtained.

\subsection{The relativistic Thomas-Fermi-Dirac model for compressed atoms}\label{sec:4ter}

We now take into account the exchange corrections to the relativistic Thomas-Fermi equation (\ref{eqless}). In this case we have (see \cite{migdal77} for instance)
\begin{equation}\label{eq:TFcondexch}
E^F_e  = \sqrt{(c P^F_e)^2+m^2_e c^4}-m_e c^2 - eV - \frac{\alpha}{\pi} c P^F_e = {\rm constant}\, .
\end{equation}

Introducing the function $\chi(r)$ as before
\begin{equation}
E^F_e + eV = e \hat{V} = \hbar c \frac{\chi}{r}\, ,
\end{equation}
we obtain the electron number density
\begin{equation}\label{eq:fullTF}
n_e = \frac{1}{3 \pi^2 \hbar^3 c^3} \left\{ \gamma \left(m_e c^2 + e \hat{V} \right)+ \left[\left(e \hat{V}\right)^2+2 m_e c^2 e \hat{V} \right]^{1/2} \left[ \frac{(1+\gamma^2) (m_e c^2 + e \hat{V})^2-m^2_e c^4}{(m_e c^2 + e \hat{V})^2-m^2_e c^4}\right]^{1/2} \right\}^3\, ,
\end{equation}
where $\gamma = (\alpha/\pi)/(1-\alpha^2/\pi^2)$.

If we take the approximation $1+\gamma^2 \approx 1$ the above equation becomes
\begin{equation}\label{eq:appTF}
n_e = \frac{1}{3 \pi^2 \hbar^3 c^3} \left\{ \gamma \left(m_e c^2 + e \hat{V} \right)+ \left[\left(e \hat{V}\right)^2 + 2 m_e c^2 e \hat{V} \right]^{1/2} \right\}^3\, .
\end{equation}
The second term on the right-hand-side of Eq.~(\ref{eq:appTF}) has the same form of the electron density given by the relativistic Thomas-Fermi approach without the exchange correction (\ref{elnd}) and therefore the first term shows the explicit contribution of the exchange term to the electron density.

Using the full expression of the electron density given by Eq.~(\ref{eq:fullTF}) we obtain the relativistic Thomas-Fermi-Dirac equation 
\begin{eqnarray}\label{eq:fullTFD}
\frac{1}{3x} \frac {d^2\chi(x)}{d x^2} &=& -\frac{\alpha}{\Delta^3}\theta( x_c- x) \nonumber \\
&+& \frac {4\alpha}{9\pi} \left\{ \gamma \left(\frac{m_e}{m_\pi}+\frac{\chi}{x}\right)+ \left[\left(\frac{\chi}{x}\right)^2+2 \frac{m_e}{m_\pi} \frac{\chi}{x}\right]^{1/2} \left[ \frac{(1+\gamma^2) (m_e/m_\pi + \chi/x)^2-(m_e/m_\pi)^2}{(m_e/m_\pi + \chi/x)^2-(m_e/m_\pi)^2}\right]^{1/2} \right\}^3 \, ,
\end{eqnarray}
which by applying the approximation $1+\gamma^2 \approx 1$ becomes
\begin{equation}\label{eq:appTFD}
\frac{1}{3x} \frac {d^2\chi(x)}{d x^2} = -\frac{\alpha}{\Delta^3}\theta( x_c- x) + \frac {4\alpha}{9\pi} \left\{ \gamma \left(\frac{m_e}{m_\pi}+\frac{\chi}{x}\right)+ \left[\left(\frac{\chi}{x}\right)^2+2 \frac{m_e}{m_\pi} \frac{\chi}{x}\right]^{1/2} \right\}^3 \, .
\end{equation}

The boundary conditions for Eq.~(\ref{eq:fullTFD}) are $\chi(0)=0$ and $\chi(x_{WS})=x_{WS}\chi'(x_{WS})$. The neutron density can be obtained as before by using the beta equilibrium condition (\ref{npeq1e}) with the electron Fermi energy given by Eq.~(\ref{eq:TFcondexch}).

\begin{figure}[t]
\begin{center}
\includegraphics[scale=0.8]{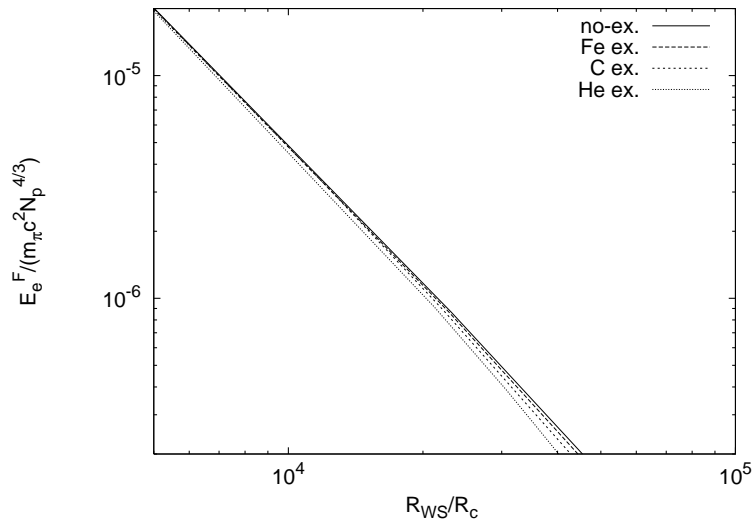}
\end{center}
\caption{The electron Fermi energies in units of $m_\pi c^2 N^{4/3}_p $ is plotted for helium, for carbon and for iron, as a function of the ratio $R_{WS}/R_c$ in the relativistic Feynman-Metropolis-Teller (FMT) treatment with and without the Thomas-Fermi-Dirac exchange effects. Here $R_{WS}$ denotes the Wigner-Seitz cell radius and $R_c$ is the nucleus radius as given by Eq.~(\ref{pn}). It is clear that the exchange terms are appreciable only in the low density region and are negligible as $R_{WS} \to R_c$}.
\label{fig:fmta}
\end{figure}

In Fig.~\ref{fig:fmta} we show the results of the numerical integration of the relativistic Thomas-Fermi equation (\ref{eqless}) and of the relativistic Thomas-Fermi-Dirac equation (\ref{eq:fullTFD}) for helium, carbon and iron. In particular, we show the electron Fermi energy multiplied by $N^{-4/3}_p$ as a function of the ratio $R_{WS}/R_c$ between the Wigner-Seitz cell radius $R_{WS}$ and the nucleus radius $R_c$ given by Eq.~(\ref{pn}).

The effects of the exchange term are appreciable only in the low density (low compression) region, i.e. when $R_{WS}>>R_c$ (see Fig.~\ref{fig:fmta}). We can then conclude in total generality that the correction given by the Thomas-Fermi-Dirac exchange term is, small in the non-relativistic low compression (low density) regime, and  negligible in the relativistic high compression (high density) regime.

\section{Comparison and contrast with approximate treatments}\label{sec:4}

There exists in the literature a variety of semi-qualitative approximations adopted in order to describe the electron component of a compressed atom (see e.g. B\"urvenich, Mishustin and Greiner \cite{mishustin} for applications of the uniform approximation and e.g Chabrier and Potekhin \cite{gilles1}, Potekhin, Chabrier and Rogers \cite {gilles2}), Haensel and Zdunik \cite{haensel1}, \cite{haensel2}, \cite{haensel4}, Douchin and Haensel \cite {haensel3} for applications of the Salpeter approximate treatment). 

We shall see how the relativistic treatment of the Thomas-Fermi equation affects  the current analysis of compressed atoms in the literature by introducing  qualitative and quantitative differences which deserve  attention.

\subsection{Relativistic FMT treatment vs. relativistic uniform approximation}

One of the most used approximations in the treatment of the electron distribution in compressed atoms is the one in which, for a given nuclear charge $+e N_p$, the Wigner-Seitz cell radius $R_{WS}$ is defined by
\begin{equation}
N_p=\frac{4 \pi}{3}R_{WS}^3n_e,
\label{m1}
\end{equation}
where $n_e=(P_e^F)^3/(3 \pi^2 \hbar^3)$. The Eq.~(\ref{m1}) ensures the global neutrality of the Wigner-Seitz cell of radius $R_{WS}$ assuming a uniform distribution of electrons inside the cell.

We shall  first compare the Feynman-Metropolis-Teller treatment, previously introduced, with the uniform approximation for the electron distribution. In view of the results of the preceding section, hereafter we shall consider the non-relativistic and the relativistic formulation of the Feynman-Metropolis-Teller treatment with no Thomas-Fermi-Dirac exchange correction.

\begin{figure}[t]
\begin{center}
\includegraphics[scale=0.65]{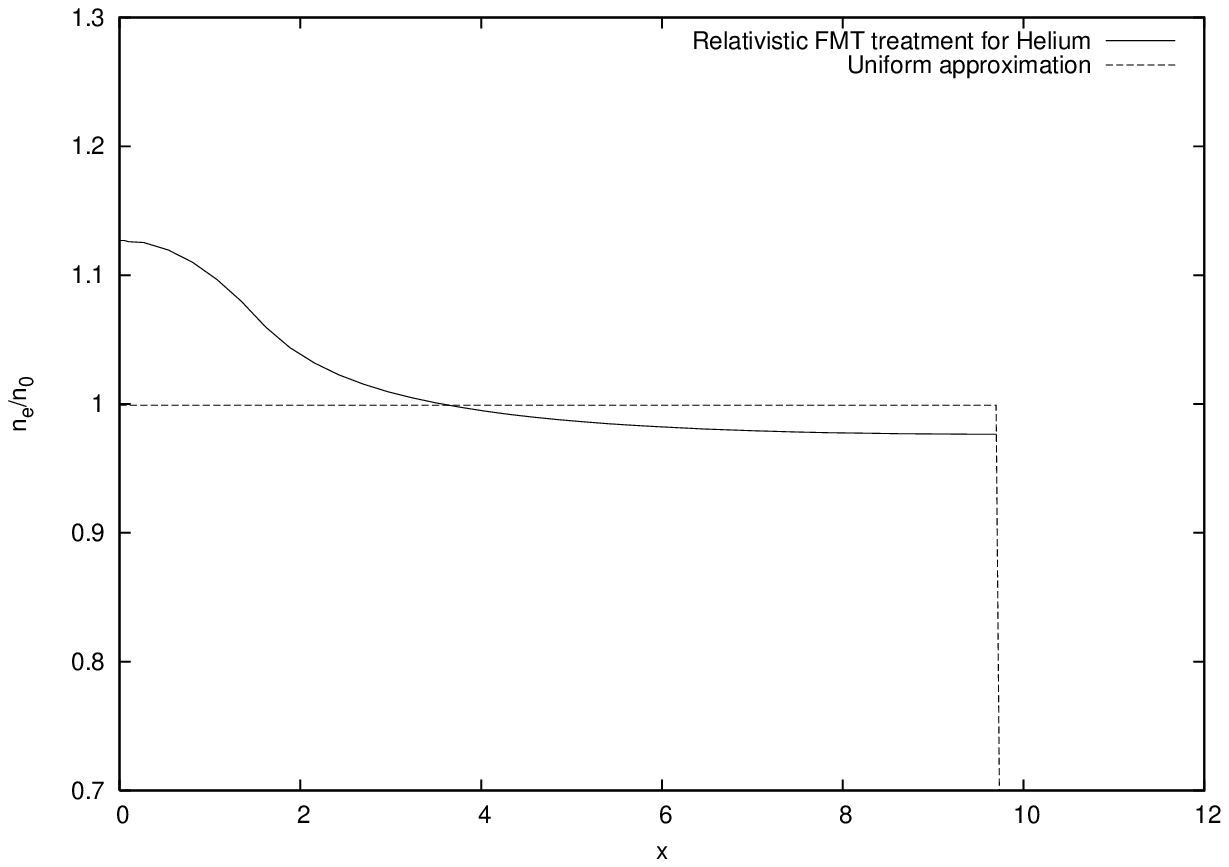}
\includegraphics[scale=0.65]{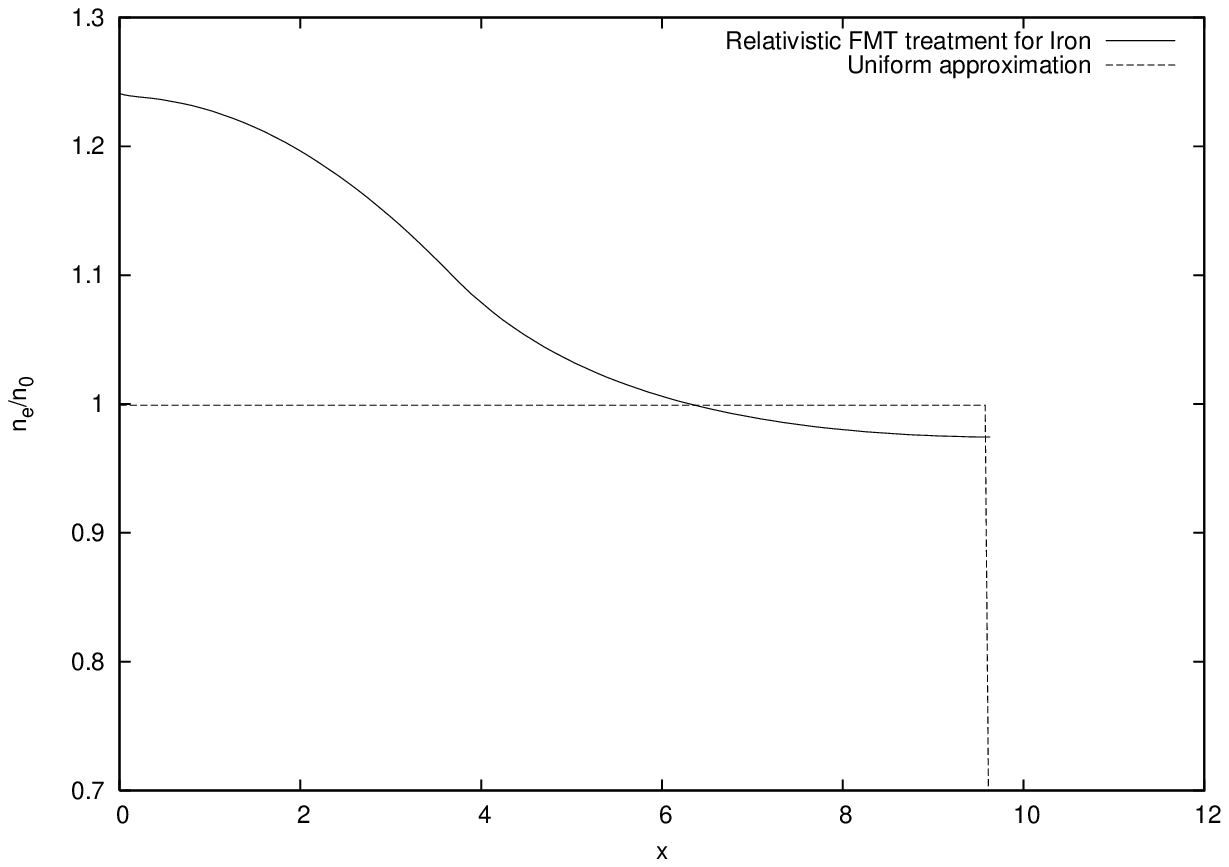}
\includegraphics[scale=0.65]{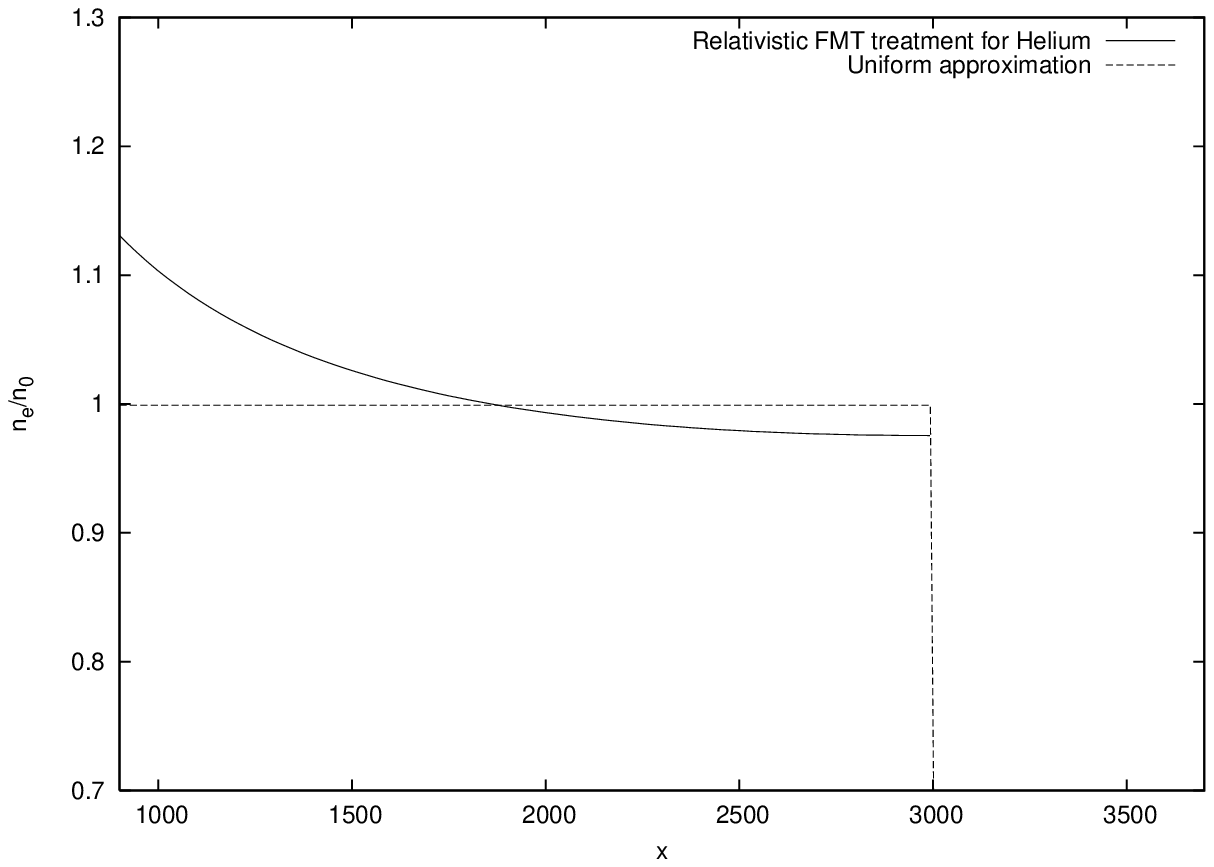}
\includegraphics[scale=0.65]{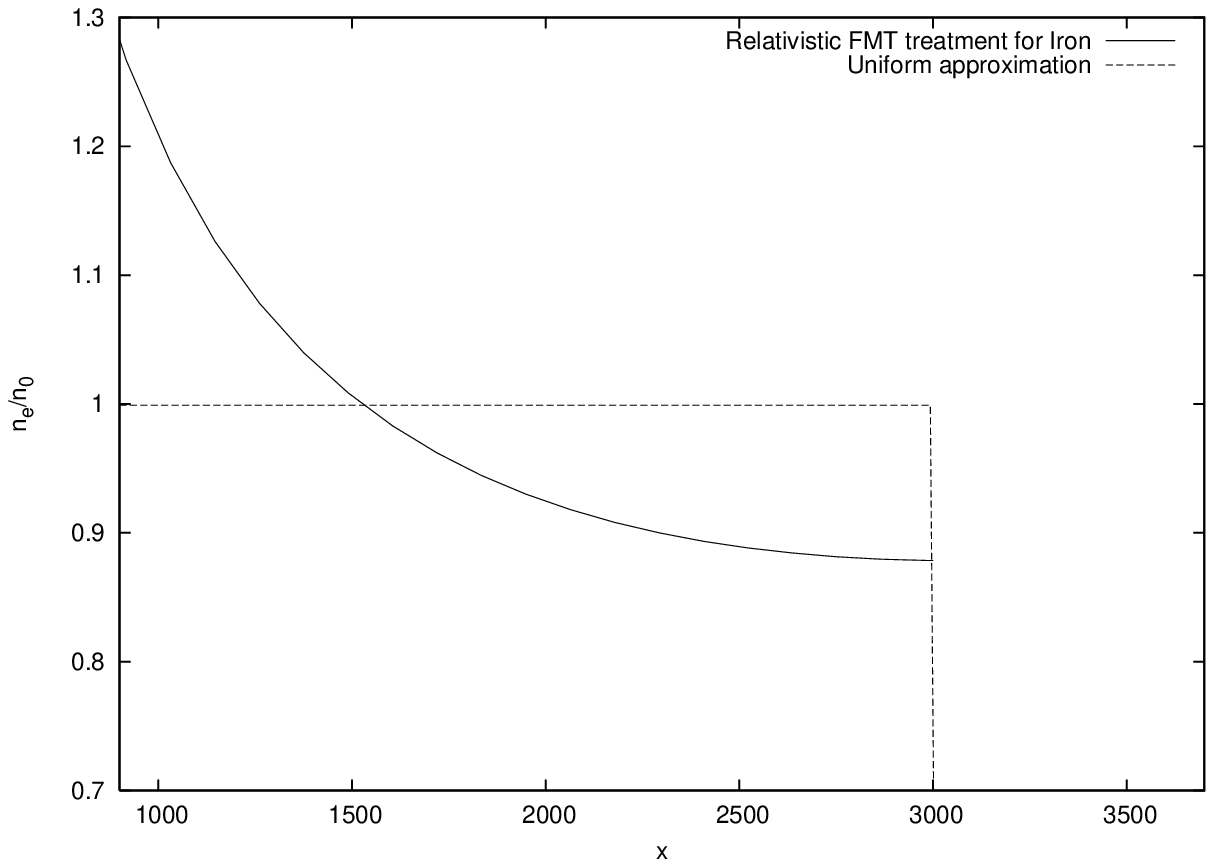}
\includegraphics[scale=0.65]{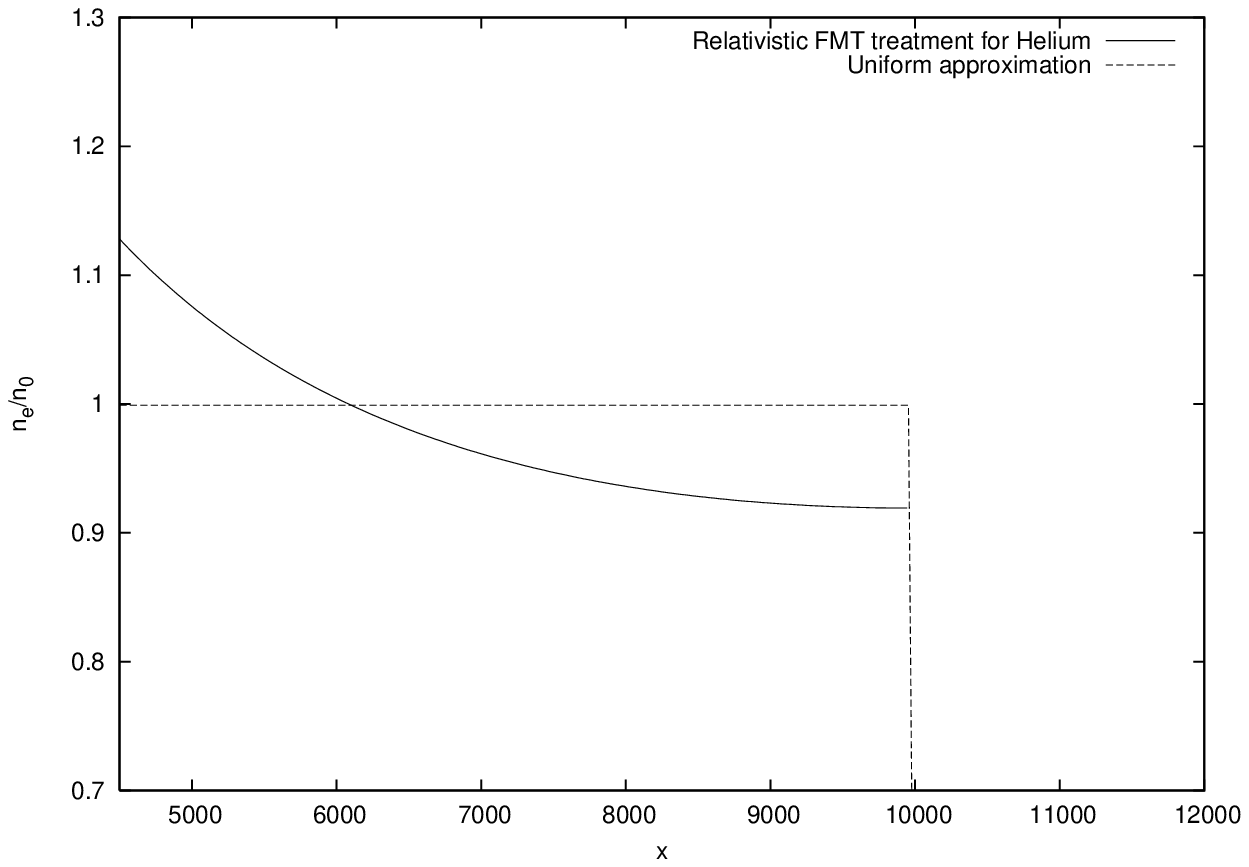}
\includegraphics[scale=0.65]{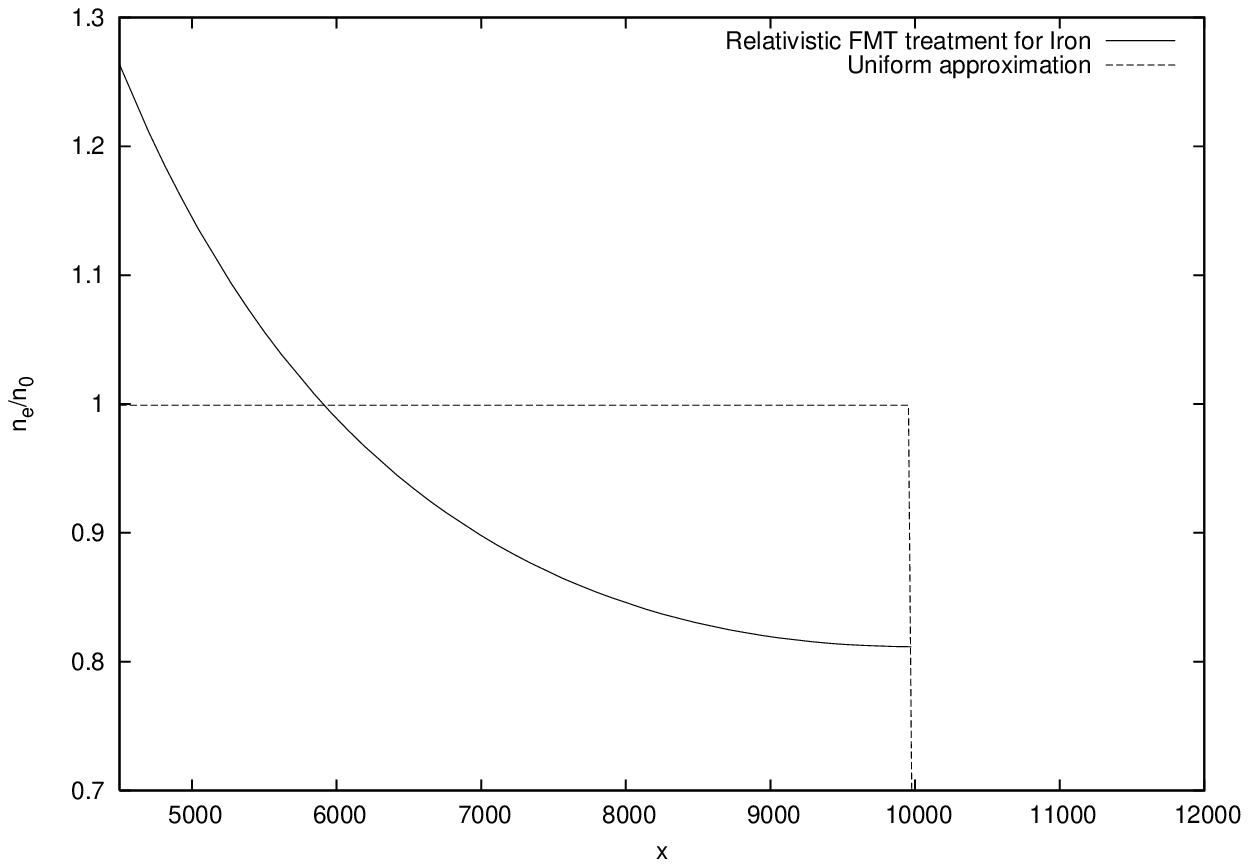}
\end{center}
\caption{The electron number density $n_e$ in units of the average electron number density $n_0 = 3 N_e/(4 \pi R_{WS}^3)$ is plotted as a function of the dimensionless radial coordinate $x=r/\lambda_\pi$ for the selected compressions $x_{WS}=9.7$ (upper panels), $x_{WS}=3\times 10^3$ (middle panels) and $x_{WS}=10^4$ (bottom panels), in both the relativistic Feynman, Metropolis, Teller approach and the uniform approximation respectively for Helium (panels on the left) and Iron (panels on the right).}
\label{fmtw}
\end{figure}

In Fig.~\ref{fmtw} we have plotted the electron number density obtained from Eq.~(\ref{elnd}) where the Coulomb potential is related to the function $\chi$, which is obtained from numerical integration of the relativistic Thomas-Fermi equation (\ref{eqless}) for different compressions for helium and iron. We have normalized the electron density to the average electron number density $n_0 = 3 N_e/(4 \pi R_{WS}^3) = 3 N_p/(4 \pi R_{WS}^3)$ as given by Eq.~(\ref{m1}).

We can see in Fig.~\ref{fmtw} how our treatment, based on the numerical integration of the relativistic Thomas-Fermi equation (\ref{eqless}) and imposing the condition of beta equilibrium (\ref{npeq1e}), leads to electron density distributions markedly different from the constant electron density approximation. 

From Eqs.~(\ref{efe}), (\ref{m1}) and taking into account the global neutrality condition of the Wigner-Seitz cell $eV(R_{WS}) = 0$, the electron Fermi energy in the uniform approximation can be written as
\begin{eqnarray}
E_e^F\simeq\left[-\frac{m_e}{m_{\pi}}+\sqrt{\left(\frac{m_e}{m_{\pi}}\right)^2+\left(\frac{9 \pi}{4}\right)^{2/3}\frac{N_p^{2/3}}{x_{WS}^2}}\right]m_{\pi}c^2.
\label{m2}
\end{eqnarray}

\begin{figure}[th]
\begin{center}
\includegraphics[scale=0.8]{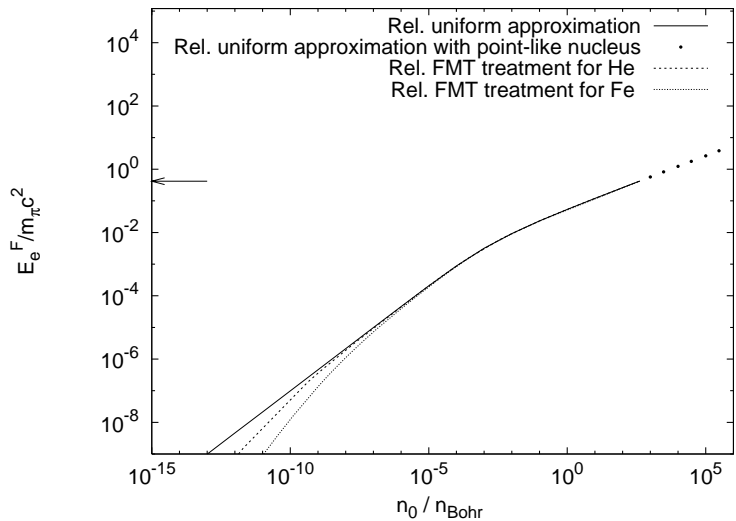}
\end{center}
\caption{The electron Fermi energies $E_e^F$ in the relativistic Feynman-Metropolis-Teller (FMT) treatment and in the uniform approximation in units of the pion rest mass, are plotted as a function of the average electron density $n_0 = 3 N_e/(4 \pi R^3_{WS})$ in units of the Bohr density $n_{\rm Bohr} = 3/(4 \pi R^3_{\rm Bohr})$ where $R_{\rm Bohr} = \hbar^2/(e^2 m_e )$ is the Bohr radius. The filled circles correspond to the case of a relativistic uniform approximation with a point-like nucleus. In such a case the electron Fermi energy can reach arbitrary large values as $R_{WS} \to 0$. The arrow indicates the value of the maximum electron Fermi energy as given by Eq.~(\ref{tf71}).}%
\label{cap1bcqs1}%
\end{figure}

We show in Fig.~\ref{cap1bcqs1} the electron Fermi energy as a function of the average electron density $n_0 = 3 N_e/(4 \pi R^3_{WS}) = 3 N_p/(4 \pi R^3_{WS})$ in units of the Bohr density $n_{\rm Bohr} = 3/(4 \pi R^3_{\rm Bohr})$ where $R_{\rm Bohr} = \hbar^2/(e^2 m_e )$ is the Bohr radius. For selected compositions we show the results for the relativistic Feynman-Metropolis-Teller treatment, based on the numerical integration of the relativistic Thomas-Fermi equation (\ref{eqless}), and for the relativistic uniform approximation. 

As clearly shown in Fig.~\ref{fmtw} and summarized in Fig.~\ref{cap1bcqs1} the relativistic treatment leads to results strongly dependent at low compression from the nuclear composition. The corresponding value of the electron Fermi energy derived from a uniform approximation overevaluates the true electron Fermi energy (see Fig.~\ref{cap1bcqs1}). In the limit of high compression the relativistic curves asymptotically approach the uniform one (see also Fig.~\ref{fmtw}). 

The uniform approximation becomes exact in the limit when the electron Fermi energy acquires its maximum value as given by 
\begin{eqnarray}
(E_e^F)_{max} \simeq \left[-\frac{m_e}{m_{\pi}}+\sqrt{\left(\frac{m_e}{m_{\pi}}\right)^2+\left(\frac{3 \pi^2}{2}\right)^{2/3}\left(\frac{N_p}{A}\right)^{2/3}}\right]m_{\pi}c^2,
\label{tf71}
\end{eqnarray} 
which is attained when $R_{WS}$ coincides with the nuclear radius $R_c$. Here, the maximum electron Fermi energy (\ref{tf71}) is obtained by replacing in Eq.~(\ref{m2}) the value of the normalized Wigner-Seitz cell radius $x_{WS} = x_c = R_c/\lambda_\pi \approx [(3/2)\pi]^{1/3}A^{1/3}$, where we have approximated the nuclear density as $n_{\rm nuc} \approx (1/2) \lambda^{-3}_\pi$.

\subsection{Relativistic FMT treatment vs. Salpeter approximate treatment}

Corrections to the uniform distribution were also studied by Salpeter \cite{salpeter} and his approximations are largely applied in physics (see e.g. Chabrier and Potekhin \cite{gilles1}, Potekhin, Chabrier and Rogers \cite {gilles2}) and astrophysics (see e.g. Haensel and Zdunik \cite{haensel1}, \cite{haensel2}, \cite{haensel4}, Douchin and Haensel \cite {haensel3} ).

Keeping the point-like nucleus assumption, Salpeter \cite{salpeter} studied the corrections to the above models due to the inhomogeneity of the electron distribution inside the Wigner-Seitz cell. He expressed an analytic formula for the total energy of a Wigner-Seitz cell  based on Coulomb corrections to the uniform distribution of electrons. The first correction   corresponds to the inclusion of the lattice energy $E_C=-(9N_p^2 \alpha)/(10 R_{WS})$, which results from the point-like nucleus-electron interaction and, from the electron-electron interaction inside the cell of radius $R_{\rm ws}$. The second correction is given by a series-expansion of the electron Fermi energy about the average electron density $n_e$ given by Eq. (\ref{m1}) the uniform approximation  $n_e = 3Z/(4 \pi R^3_{\rm ws})$. The electron density is then assumed equals to $n_e [1+\epsilon(r)]$ with $\epsilon(r)$ considered as infinitesimal. The Coulomb potential energy is assumed to be the one of the point-like nucleus with the uniform distribution of electrons of density $n_e$ given by, thus the correction given by $\epsilon(r)$ is neglected on the Coulomb potential. The electron distribution is then calculated at first-order by expanding the relativistic electron kinetic energy 
\begin{eqnarray}
\epsilon_k &=& \sqrt{[c P^F_e(r)]^2 + m^2_e c^4}-m_e c^2 \nonumber \\
&=& \sqrt{(3 \pi^2 n_e)^{2/3}[1+\epsilon(r)]^{2/3} + m^2_e c^4}-m_e c^2 ,
\end{eqnarray}
about its value given by the uniform approximation 
\begin{equation}
\epsilon^{\rm unif}_k = \sqrt{(3 \pi^2 n_e)^{2/3} + m^2_e c^4}-m_e c^2\, ,
\end{equation}
considering as infinitesimal the ratio $eV/E^F_e$ between the Coulomb potential energy $eV$ and the electron Fermi energy $E^F_e = \sqrt{[c P^F_e(r)]^2 + m^2_e c^4}-m_e c^2 - e V$.

The effect of the Dirac electron-exchange correction \cite{dirac30} on the equation of state was also considered by Salpeter \cite{salpeter}. However,  adopting the general approach of Migdal et al. \cite{migdal77}, these effects are negligible in the relativistic regime (see Subsec. \ref{sec:4ter} ).

The inclusion of each additional Coulomb correction results in a decreasing of the pressure of the cell $P_{S}$ (see \cite{inpreparation}  for details). However, despite to be very interesting in identifying piecewise contributions  to the total pressure,  the validity of the Salpeter approach needs a verification by a more general treatment. For instance, the failure of the Salpeter formulas can be  seen at  densities of the order of $\sim 10^2-10^3$ g cm$^{-3}$  for nuclei with large $N_p$, as in the case of iron,  where the pressure becomes negative (see Table (\ref{xx1})). Therefore, the problem of solving the relativistic Thomas-Fermi equation within the Feynman, Metropolis, Teller approach  becomes a necessity, since this approach gives  all the possible Coulomb and relativistic contributions automatically and correctly.

\begin{table}
\caption{Pressure for Iron as a function of the density $\rho$ in the uniform approximation ($P$), in the  Salpeter approximation ($P_S$) and in the relativistic Feynman-Metropolis-Teller approach ($P_{FMTrel}$). Here $x_{S}=P_{e,S}^F/(m_ec)$, $x_{FMTrel}=P_e^F/(m_ec)$ are respectively the normalized Salpeter Fermi momentum and the relativistic Feynmann-Metropolis-Teller Fermi momentum.}
\label{xx1}
\begin{center}
\begin{tabular}{|c|c|c|c|c|c|}
	\hline $    \rho  $ &  $ x_{S}$ &  $ x_{FMTrel} $ & $  P            $& $ P_S $ & $ P_{FMTrel} $ \\   
	       $    ({\rm g/cm}^{3})      $ &  $              $ & $ $ & $ ({\rm bar})   $ &$  ({\rm bar})   $ & $ ({\rm bar})$ \\ 
  \hline $      2.63\times 10^2   $ &      $0.05      $ &     $ 0.0400  $  &  $2.9907\times10^{10}$& $-1.8800\times10^{8}   $ &  $  9.9100\times10^{9}    $            \\ 
	       $  2.10\times10^3       $ &      $0.10      $ &     $0.0857   $  & $9.5458\times10^{11}$ &  $ 4.4590\times10^{11}  $ &  $5.4840\times10^{11}      $            \\ 
	       $  1.68\times10^4       $ &      $0.20      $ &     $0.1893   $  & $ 3.0227\times10^{13}$ &  $2.2090\times10^{13}   $ &  $2.2971\times10^{13}      $             \\ 
	       $ 5.66\times10^4        $ &      $0.30      $ &     $0.2888   $  & $2.2568\times10^{14}$ &  $1.8456\times10^{14}   $ &  $1.8710\times10^{14}      $             \\ 
	       $ 1.35\times10^5        $ &      $0.40      $ &     $0.3887   $  & $9.2964\times10^{14}$ &  $8.0010\times10^{14}   $ &  $8.0790\times10^{14}      $              \\
	       $  2.63\times10^5       $ &      $0.50      $ &     $0.4876   $  &  $2.7598\times10^{15}$ & $2.4400\times10^{15}   $ &  $2.4400\times10^{15}      $                \\  
	       $ 4.53\times10^5        $ &      $0.60      $ &     $0.5921   $  & $6.6536\times10^{15}$ &  $6.0040\times10^{15}   $ &  $6.0678\times10^{15}      $            \\ 
	       $  7.19\times10^5       $ &      $0.70      $ &     $0.6820   $  &  $1.3890\times10^{16}$ & $1.2693\times10^{16}   $ &  $1.2810\times10^{16}      $            \\ 
	       $ 1.08\times10^6        $ &      $0.80      $ &     $0.7888   $  & $ 2.6097\times10^{16}$ &  $2.4060\times10^{16}   $ &  $2.4442\times10^{16}      $             \\ 
	       $ 2.10\times10^6        $ &      $1.00      $ &     $0.9853   $  &  $7.3639\times10^{16}$ & $6.8647\times10^{16}   $ &  $6.8786\times10^{16}      $             \\ 
	       $ 3.63\times10^6        $ &      $1.20      $ &     $1.1833   $  & $1.6902\times10^{17}$ &  $1.5900\times10^{17}   $ &  $1.5900\times10^{17}      $              \\
	       $   5.77\times10^6      $ &      $1.40      $ &     $1.3827   $  & $3.3708\times10^{17}$ &  $3.1844\times10^{17}   $ &  $3.1898\times10^{17}        $ \\     $8.62\times10^6         $ &      $1.6      $ &     $1.5810   $  & $6.0754\times10^{17}$ &  $5.7588\times10^{17}   $ &  $5.7620\times10^{17}      $              \\
	       $1.23\times10^7         $ &      $1.80      $ &     $ 1.7790  $  & $1.0148\times10^{18}$ &  $9.6522\times10^{17}   $ &  $9.6592\times10^{17}      $                \\
	       $1.68\times10^7         $ &      $2.0     $ &     $1.9770   $  &  $ 1.5981\times10^{18}$ & $1.5213\times10^{18}   $ &  $1.5182\times10^{18}      $              \\
	       $3.27\times10^7         $ &      $2.50      $ &     $2.4670   $  &  $4.1247\times10^{18}$ & $ 3.9375\times10^{18}  $ &  $3.9101\times10^{18}      $                \\
	       $5.66\times10^7         $ &      $3.00      $ &     $2.965   $  &  $8.8468\times10^{18}$ & $8.4593\times10^{18}   $ &  $8.4262\times10^{18}      $              \\
	       $ 1.35\times10^8        $ &      $4.00      $ &     $3.956   $  &  $2.9013\times10^{19}$ & $2.7829\times10^{19}   $ &  $2.7764\times10^{19}      $                \\
	       $  2.63\times10^8       $ &      $5.00     $ &     $4.939   $  &  $7.2160\times10^{19}$ & $6.9166\times10^{19}   $ &  $6.9062\times10^{19}      $              \\
	       $  8.85\times10^8       $ &      $7.50      $ &     $7.423   $  & $3.7254\times10^{20}$ &  $3.5700\times10^{20}   $ &  $3.5700\times10^{20}      $                \\     
	       \hline	
\end{tabular}
\end{center}
\end{table}

\subsection{Relativistic FMT treatment vs. non-relativistic FMT treatment}

In order to compare and contrast the Fermi energy of a compressed atom in the non-relativistic and the relativistic  limit we first express the non-relativistic equations in terms of the dimensionless variables used for the relativistic treatment.
We then have
\begin{eqnarray}
x=\frac{r}{\lambda_{\pi}}, \quad \frac{\chi}{r}=\frac{e \hat V}{c\hbar},
\label{tf4}
\end{eqnarray} 
and 
the non-relativistic limit of Eq.~($\ref{eqless}$) becomes
\begin{eqnarray}
\frac {d^2\chi(x)}{d x^2}=\frac{2^{7/2}}{3 \pi}\alpha\left(\frac{m_e}{m_{\pi}}\right)^{3/2}\frac {\chi^{3/2}}{ x^{1/2}} ,
\label{tf5}
\end{eqnarray} 
with the boundary conditions
\begin{eqnarray}
\chi(0)=\alpha N_p, \quad x_{WS}\chi(x_{WS})'=\chi(x_{WS}),
\label{tf6}
\end{eqnarray}
and dimensionless variable $x_{WS}=R_{WS}/\lambda_{\pi}$.

In these new variables the electron Fermi energy is given by
\begin{eqnarray}
E_e^F=\frac{\chi(x_{WS})}{x_{WS}}m_{\pi}c^2.
\label{tf7}
\end{eqnarray} 

\begin{figure}[ht]
\begin{center}
\includegraphics[scale=0.8]{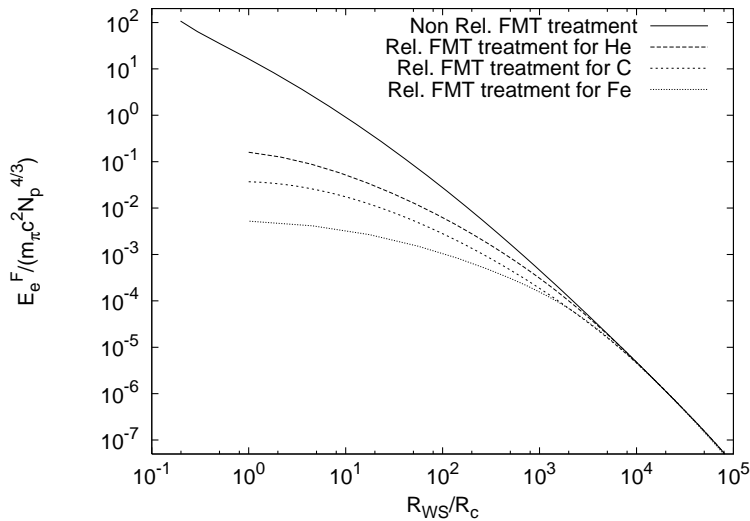}
\end{center}
\caption{The electron Fermi energies in units of $m_\pi c^2 N^{4/3}_p $ for helium, for carbon and for iron are plotted as a function of the ratio $R_{WS}/R_c$ respectively in the non-relativistic and in the relativistic Feynman-Metropolis-Teller (FMT) treatment without the Thomas-Fermi-Dirac exchange effects. Here $R_{WS}$ is the radius of the Wigner-Seitz cell and $R_c$ is the radius of the nucleus given by Eq.~(\ref{pn}). The relativistic treatment leads to results of the electron Fermi energy strongly dependent on the nuclear composition and systematically smaller than the non-relativistic ones, which can attain arbitrary large values as the point-like nucleus is approached.}
\label{fmt}
\end{figure}

The two treatment, the relativistic and the non-relativistic one, can be now directly compared and contrasted by using the same units (see Fig.~\ref{fmt}).

There are major differences:

1) \quad The electron Fermi energy in the relativistic treatment is strongly dependent on the nuclear composition, while the non-relativistic treatment presents a universal behavior in the units of Figs.~\ref{fmt}. In the limit of low densities the relativistic curves approach the universal non-relativistic curve.

2) \quad  The relativistic treatment leads to values of the electron Fermi energy consistently smaller than the ones of the non-relativistic treatment. 

3) \quad While in the non-relativistic treatment the electron Fermi energy can reach, by compression, infinite values as $R_{WS}\rightarrow0$, in the relativistic treatment it reaches a perfectly finite value given by Eq.~(\ref{tf71}) attained when $R_{WS}$ coincides with the nuclear radius $R_c$.

The universality of the electron Fermi energy with respect to the number of protons $N_p$ has been obtained by expressing the Coulomb potential energy $e V$ in terms of the function $\phi$ given by Eq.~(\ref{ecn5}), and by introducing the scale factor $b$ given by (\ref{ecn8}). Accordingly, the radius of the Wigner-Seitz cell has been expressed in terms of the nucleus radius (\ref{pn}) which is proportional to $N^{1/3}_p$.

It is clear then, from above considerations, the relativistic treatment of the Thomas-Fermi equation introduces significant differences from the current approximations in the literature: a) the uniform electron distribution \cite{mishustin}, b) the approximate perturbative solutions departing from the uniform distribution \cite{salpeter} and c) the non-relativistic treatment \cite{fmt}.
We have recently applied these results of the relativistic Feynman, Metropolis, Teller treatment of a compressed atom to the study of white dwarfs and their consequences on the determination of their masses, radii and critical mass \cite{inpreparation}.

\section{Application to nuclear matter  cores of stellar dimensions}\label{sec:5}

We turn now to nuclear matter cores of stellar dimensions of $A\simeq(m_{\rm Planck}/m_n)^3 \sim 10^{57}$ or $M_{core} \sim M_{\odot}$.

Following the treatment presented in Popov et al.,\cite{prl}, we use the existence of  scaling laws and proceed to the ultra-relativistic limit of Eqs. (\ref{pnx}), (\ref{elnd}), (\ref{eqless}), (\ref{npeq1e}). For positive values of the electron Fermi energy $E_e^F$, we introduce the new function  $\phi=4^{1/3}(9\pi)^{-1/3} \chi \Delta/x$ and the new variable $\hat x=kx$ where $k=\left(12/\pi\right)^{1/6}\sqrt{\alpha}\Delta^{-1}$, as well as the variable  $\xi=\hat x- \hat x_c$ in order to describe better the region around the core radius.

Eq.~(\ref{eqless}) becomes
\begin{eqnarray}
\frac {d^2\hat \phi(\xi)}{d \xi ^2}=-\theta(-\xi)+\hat \phi(\xi)^3\,,
\label{eqless5}
\end{eqnarray}
where  $\hat \phi(\xi)=\phi(\xi+\hat x_c)$ and the curvature term $2\hat \phi'(\xi)/(\xi+\hat x_c)$ has been neglected.

The Coulomb potential energy is given by
\begin{eqnarray}
eV(\xi)=\left(\frac{9\pi}{4}\right)^{1/3}\frac{1}{\Delta} m_\pi c^2 \hat \phi(\xi)-E_e^F\,,
\label{v01}
\end{eqnarray}
corresponding to the electric field
\begin{eqnarray}
E(\xi)=-\left(\frac{3^5\pi}{4}\right)^{1/6}\frac{\sqrt{\alpha}}{\Delta^2}\frac{m_\pi^2 c^3}{e\hbar }  \hat \phi'(\xi),
\label{v01e}
\end{eqnarray}
and the electron number-density 
\begin{eqnarray}
n_e(r) = \frac {1}{3\pi^2\hbar^3c^3}
\left(\frac{9\pi}{4}\right)\frac{1}{\Delta^3} (m_\pi c^2)^3 \hat \phi^3(\xi). 
\label{elnd1}
\end{eqnarray}
In the core center we must have $n_e=n_p$. From Eqs.~(\ref{pnx}) and (\ref{elnd1}) we than have that, for $\xi=-\hat x_c$, $\hat \phi(-\hat x_c)=1$.

In order to consider a compressed nuclear density core of stellar dimensions, we then introduce a Wigner-Seitz cell determining the outer boundary of the electron distribution  which, in the new radial coordinate $\xi$ is characterized by $\xi^{WS}$. In view of the global charge neutrality of the system the electric field goes to zero at $\xi=\xi^{WS}$. This implies, from Eq. (\ref{v01e}), $\hat \phi'(\xi^{WS})=0$.

We now turn to the determination of the Fermi energy of the electrons in this compressed core. The function $\hat \phi$ and  its first derivative $\hat \phi'$ must be continuous at the surface $\xi=0$ of the nuclear density core. 

\begin{figure}[t]
\begin{center}
\includegraphics[scale=0.8]{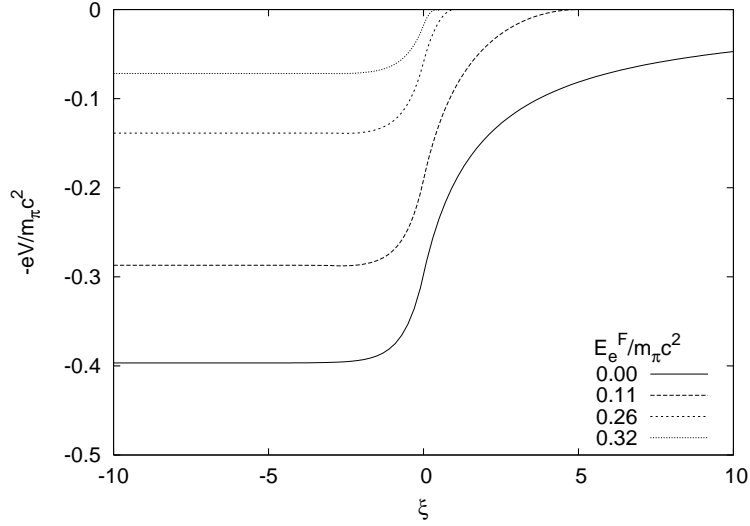}
\end{center}
\caption{The electron Coulomb potential energies in units of the pion rest mass in a nuclear matter core of stellar dimensions with $A\simeq 10^{57}$ or $M_{core} \sim M_{\odot}$ and $R_c \approx 10^6$ cm, are plotted as a function of the dimensionless variable $\xi$, for different values of the electron Fermi energy also in units of the pion rest mass. The solid line corresponds to the case of null electron Fermi energy. By increasing the value of the electron Fermi energy the electron Coulomb potential energy depth is reduced.}
\label{efieldf}
\end{figure}

\begin{figure}[t]
\begin{center}
\includegraphics[scale=0.8]{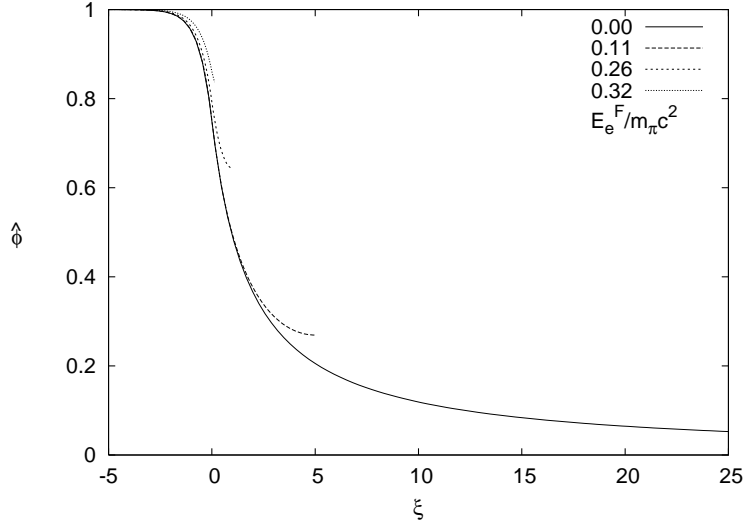}
\end{center}
\caption{Solutions of the ultra-relativistic Thomas-Fermi equation (\ref{eqless5}) for different values of the Wigner-Seitz cell radius $R_{WS}$ and correspondingly of the electron Fermi energy in units of the pion rest mass as in Fig.~\ref{efieldf}, near the core surface. The solid line corresponds to the case of null electron Fermi energy.}
\label{efieldf1}
\end{figure}

\begin{figure}[th]
\begin{center}
\includegraphics[scale=0.8]{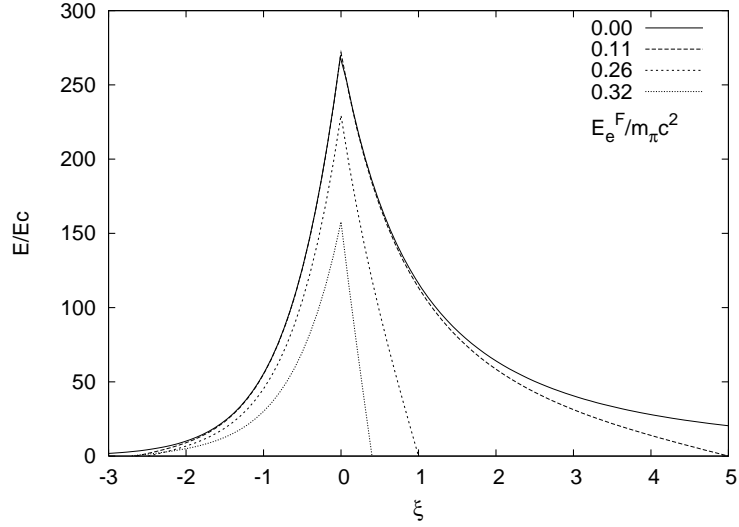}
\end{center}
\caption{The electric field in units of the critical field for vacuum polarization $E_c=m_e^2c^3/(e\hbar)$ is plotted as a function of the coordinate $\xi$, for different values of the electron Fermi energy in units of the pion mass. The solid line corresponds to the case of null electron Fermi energy. To an increase of the value of the electron Fermi energy it is found a reduction of the  peak of the electric field. 
}%
\label{ANp}
\end{figure}

\begin{figure}[th]
\begin{center}
\includegraphics[scale=0.8]{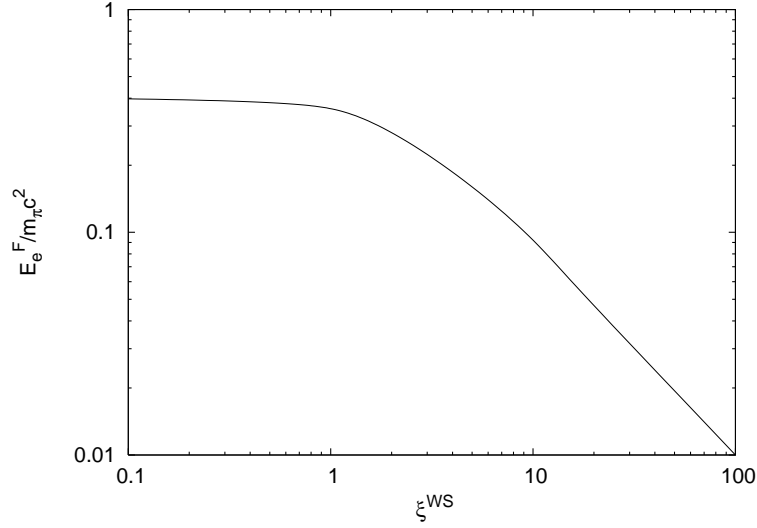}
\end{center}
\caption{The Fermi energy of electrons in units of the pion rest mass is plotted for different Wigner-Seitz cell dimensions (i.e for different compressions) $\xi^{WS}$ in the ultra-relativistic approximation . In the limit $\xi^{WS}\rightarrow 0$ the electron Fermi energy approaches asymptotically the value $(E_e^F)_{max}$ given by Eq.~(\ref{efa1}). 
}%
\label{ANp1}
\end{figure}

This boundary-value problem  can be solved analytically and indeed Eq.~(\ref{eqless5}) has the first integral,
\begin{equation}
2[\hat \phi'(\xi)]^2 = \left\{\begin{array}{ll} \hat \phi^4(\xi)-4\hat \phi(\xi) +3, &  \quad \xi<0, \\
\hat \phi^4(\xi)-\phi^4(\xi^{WS}), & \quad \xi>0,
\end{array}\right.
\label{popovs1}
\end{equation} 
with boundary conditions at $\xi=0$: 
\begin{eqnarray}
\hat \phi(0)&=&\frac{\hat \phi^4(\xi^{WS})+3}{4},\nonumber\\
\hat \phi'(0) &=&-\sqrt{\frac{\hat \phi^4(0)-\hat \phi^4(\xi^{WS})}{2}}.
\label{b1}  
\end{eqnarray}
Having fullfilled the continuity condition we integrate Eq. (\ref{popovs1}) obtaining  for $\xi\le 0$  
\begin{eqnarray}
\hat \phi(\xi) 
= 1-3\left[1+2^{-1/2}\sinh(a-\sqrt{3}\xi)\right]^{-1}, 
\label{2nd1}
\end{eqnarray}
where the integration constant $a$ has the value 
\begin{eqnarray}
{\rm sinh}(a)=\sqrt{2}\left(\frac{11+\hat \phi^4(\xi^{WS})}{1-\hat \phi^4(\xi^{WS})}\right).
\label{b11}
\end{eqnarray}
In the interval  $0\le \xi\le \xi^{WS}$, the field $\hat \phi(\xi) $  is
implicitly given by
\begin{eqnarray}
F\left({\rm arccos}\frac{\hat\phi(\xi^{WS})}{\hat\phi(\xi)},\frac{1}{\sqrt{2}}\right)
=\hat\phi(\xi^{WS})(\xi-\xi^{WS}),
\label{2nd2}
\end{eqnarray}
where $F(\varphi,k)$ is the elliptic function of the first kind, 
and $F(0,k)\equiv 0$.
For $F(\varphi,k)=u$, 
the inverse function $\varphi=F^{-1}(u,k)={\rm am}(u,k)$ is the well known Jacobi amplitude. In terms of it, we can express the solution (\ref{2nd2}) for $\xi > 0$
as,
\begin{eqnarray}
\hat\phi(\xi)&=&\hat\phi(\xi^{WS})\left\{\cos\left[{\rm am}\left(\hat\phi(\xi^{WS})(\xi-\xi^{WS}),\frac{1}{\sqrt{2}}\right)\right]\right\}^{-1}.
\label{2nd2s}
\end{eqnarray}

In the present case of $E_e^F > 0$ the ultra-relativistic approximation is indeed always valid up to $\xi=\xi^{WS}$ for high compression factors, i.e. for $R_{WS}\simeq R_c$. In the case $E_e^F=0$, $\xi^{WS}\rightarrow \infty$, there is a breakdown of the ultra-relativistic approximation when $\xi\rightarrow \xi^{WS}$.
 
Details are given in Figs. \ref{efieldf}, \ref{efieldf1}, \ref{ANp}.

We can now  estimate two crucial quantities of the solutions: the Coulomb potential at the center of the configuration and the electric field at the surface of the core
\begin{eqnarray}
eV(0)\simeq \left(\frac{9\pi}{4}\right)^{1/3} \frac{1}{\Delta} m_\pi c^2-E_e^F ,
\label{v0}
\end{eqnarray}
\begin{eqnarray}
E_{\rm max}& \simeq & 2.4\frac{\sqrt{\alpha}}{\Delta^2}\left(\frac{m_{\pi}}{m_e}\right)^2E_c
|\hat\phi'(0)|\,, 
\label{v0w}
\end{eqnarray}
where $E_c=m_e^2c^3/(e\hbar)$ is the critical electric field for vacuum polarization. These functions depend on the value $\hat\phi(\xi^{WS})$ via Eqs.~(\ref{popovs1})--(\ref{2nd2}). At the boundary $\xi=\xi^{WS}$, due to the global charge neutrality, both the electric field $E(\xi^{WS})$ and  the Coulomb potential $eV(\xi^{WS})$ vanish. From Eq.~(\ref{v01}), we determine the value of $\hat \phi(\xi)$ at $\xi=\xi^{WS}$
\begin{eqnarray}
\hat \phi(\xi^{WS})=\Delta\left(\frac{4}{9\pi}\right)^{1/3}\frac{E_e^F}{m_\pi c^2} \,,
\label{v02}
\end{eqnarray}
as a function of the electron Fermi energies $E_e^F$. From the above Eq.~(\ref{v02}), one 
can see that there exists a solution, characterized by the value of electron Fermi energy 
\begin{eqnarray}
\frac{(E_e^F)_{max}}{m_\pi c^2}=\frac{1}{\Delta}\left(\frac{9\pi}{4}\right)^{1/3} \, ,
\label{v03}
\end{eqnarray}
such that $\hat \phi(\xi^{WS})=1$. From Eq.~(\ref{2nd2}) and $\xi=0$, we also have
\begin{eqnarray} 
\xi^{WS}(\hat \phi(\xi^{WS}))=\left\{\frac{1}{\hat \phi(0)} F\left[arccos\left(4-\frac{3}{\hat \phi(0)}\right), \frac{1}{\sqrt{2}}\right]\right\}.
\label{pn2q}
\end{eqnarray}
For $\hat \phi(\xi^{WS})= 1$, from Eq.~(\ref{b1}) follows $\hat \phi(0)=1$ hence Eq.~(\ref{pn2q}) becomes
\begin{eqnarray} 
\xi^{WS}(\hat \phi(0))= F\left[0, \frac{1}{\sqrt{2}}\right].
\label{pn2}
\end{eqnarray}
It is well known that if the inverse Jacobi amplitude $F[0,1/\sqrt{2}]$ is zero, then 
\begin{eqnarray} 
\xi^{WS}(\hat \phi(\xi^{WS})=\hat \phi(0)=1)=0.
\label{pn3}
\end{eqnarray}
Indeed from $\hat \phi(\xi^{WS})=1$ follows $\hat \phi(0)=1$ and $\xi^{WS}=0$. When $\xi^{WS}=0$ from Eq. (\ref{b1}) follows $\hat \phi'(0)=0$ and, using Eq. (\ref{v0w}), $E_{\rm max }=0$. In other words for the value of $E_e^F$ fulfilling Eq. (\ref{v02}) no electric field exists on the boundary of the core and from  Eq. (\ref{elnd1}) and Eqs. (\ref{pn}, \ref{pnx}) it follows that indeed this is the solution fulfilling both global $N_e=N_p$ and local $n_e=n_p$ charge neutrality. In this special case, starting from Eq.~(\ref{npeq1e}) and $A=N_p+N_n$, we obtain
\begin{eqnarray}
(E_e^F)_{max}^{3/2}=\frac{\frac{9 \pi}{4}(\hbar c)^3\frac{A}{R_c^3}-(E_e^F)_{max}^3}{2^{3/2}\left[\left(\frac{9 \pi}{4}(\hbar c)^3 \frac{A}{R_c^3}  -(E_e^F)_{max}^{3} \right)^{2/3} +m_n^2c^4\right]^{3/4}  } .
\label{efa}
\end{eqnarray}
In the ultra-relativistic approximation $(E_e^F)^3_{max}/\frac{9 \pi}{4} (\hbar c)^3\frac{A}{R_c^3}<<1$ so Eq.~(\ref{efa}) can be approximated to
\begin{eqnarray}
(E_e^F)_{max}=2^{1/3}\frac{m_n}{m_{\pi}}\gamma\left[-1+\sqrt{1+\frac{\beta}{2\gamma^3}}\right]^{2/3}m_{\pi} c^2 ,
\label{efa1}
\end{eqnarray}
where
\begin{eqnarray}
 \beta=\frac{9 \pi}{4} \left(\frac{\hbar}{m_n c}\right)^3 \frac{A}{R_c^3}, \qquad \gamma=\sqrt{1+\beta^{2/3}} .
\label{efa2}
\end{eqnarray}

The corresponding limiting value to the $N_p/A$ ratio is obtained as follows
\begin{eqnarray}
\frac{N_p}{A}=\frac{2 \gamma^3}{\beta} \left[-1+\sqrt{1+\frac{\beta}{2\gamma^3}}\right]^{2} .
\label{efa3}
\end{eqnarray}
Inserting Eqs.~(\ref{efa1}), (\ref{efa2}) in Eq.~(\ref{efa3}) one obtains the ultra-relativistic limit of Eq.~(\ref{tf71}), since the electron Fermi energy, in view of the scaling laws introduced in \cite{prl}, is independent of the value of $A$ and depends only on the density of the core. 

The $N_p$-independence in the limiting case of maximum electron Fermi energy attained when $R_{WS}=R_c$, in which the ultra-relativistic treatment approaches the uniform one, and the $N_p$-dependence for smaller compressions $R_{WS} > R_c$ can be understood as follows. Let see the solution to the ultra-relativistic equation (\ref{eqless5}) for small $\xi>0$. Analogously to the Feynman-Metropolis-Teller approach to the non-relativistic Thomas-Fermi equation, we solve the ultra-relativistic equation (\ref{eqless5}) for small $\xi$. 
Expanding $\hat\phi(\xi)$ about $\xi=0$ in a semi convergent power series, 
\begin{eqnarray}
\frac{\hat\phi(\xi)}{\hat\phi(0)}=1+\sum_{n=2}^\infty a_n\xi^{n/2}
\label{exp}
\end{eqnarray}
and substituting it into the ultra-relativistic equation (\ref{eqless5}), we have
\begin{eqnarray}
\sum_{k=3}^\infty a_k\frac{k(k-2)}{4}\xi^{(k-4)/2}=\phi^2(0)\exp \left[3\ln (1+\sum_{n=2}^\infty a_n\xi^{n/2})\right].
\label{expeq}
\end{eqnarray}
This leads to a recursive determination of the coefficients:
\begin{eqnarray}
a_3=0,\, a_4=\phi^2(0)/2,\, a_5=0,\, a_6=\phi^2(0)a_2/2,\,a_7=0,\,
a_8=\phi^2(0)(1-a_2^2)/8,\cdot\cdot\cdot,
\label{expan}
\end{eqnarray}
with $a_2=\hat\phi'(0)/\hat\phi(0)$ 
determined by the initial slop, namely, the boundary condition $\hat\phi'(0)$ and $\hat\phi(0)$ in Eq.~(\ref{b1}):
\begin{eqnarray}
\hat\phi(0)=\frac{\hat\phi^4(\xi^{WS})+3}{4},\quad \hat\phi'(0)=-\sqrt{\frac{\hat\phi^4(0)-\hat\phi^4(\xi^{WS})}{2}}
\label{bcon}
\end{eqnarray}
Thus the series solution (\ref{exp}) is uniquely determined by the boundary value $\hat\phi(\xi^{WS})$ at the Wigner-Seitz cell radius. 

Now we consider the solution up to the leading orders
\begin{eqnarray}
\hat\phi(\xi)=\hat\phi(0)+\hat\phi'(0)\xi + \frac{1}{2}\hat\phi^3(0)\xi^2
+\frac{1}{2}\hat\phi^3(0)a_2\xi^3+\frac{1}{8}\hat\phi^3(0)(1-a_2^2)\xi^4
+\cdot\cdot\cdot.
\label{exp1}
\end{eqnarray}
Using Eq.~(\ref{exp1}), the electron Fermi energy (\ref{v02}) becomes
\begin{equation}
E^F_e= (E^F_e)_{max} \left[1+a_2\xi^{WS}+\frac{1}{2}\hat\phi^2(0)(\xi^{WS})^2+\frac{1}{2}\hat\phi^2(0)a_2(\xi^{WS})^3+\frac{1}{8}\hat\phi^2(0)(1-a_2^2)(\xi^{WS})^4
+\cdot\cdot\cdot\right] \hat\phi(0),
\label{tf_corr_fermi}
\end{equation}
where $(E^F_e)_{max}=(9\pi/4)^{1/3}\Delta^{-1}$ is the maximum Fermi energy which is attained when the Wigner-Seitz cell radius equals the nucleus radius $R_c$ (see Eq.~\ref{v03}). For $\hat\phi(\xi^{WS})<1$, we approximately have $\hat\phi(0)=3/4$, $\hat\phi'(0)=-(3/4)^2/\sqrt{2}$ and the initial slope $a_2=\hat\phi'(0)/\hat\phi(0)=-(3/4)/\sqrt{2}$. Therefore Eq.~(\ref{tf_corr_fermi}) becomes
\begin{equation}
E^F_e\approx (E^F_e)_{max} \left[1-\frac{3}{4\sqrt{2}}\xi^{WS}+\frac{1}{2}\left(\frac{3}{4}\right)^2(\xi^{WS})^2-\frac{1}{2^{3/2}}\left(\frac{3}{4}\right)^3(\xi^{WS})^3+\frac{1}{8}\left(\frac{3}{4}\right)^2\left(\frac{41}{32}\right)(\xi^{WS})^4+\cdot\cdot\cdot\right].
\label{tf_corr_fermi_1}
\end{equation}

By the definition of the coordinate $\xi$, we know all terms except the first term in the square bracket depend on the values of $N_p$. In the limit of maximum compression when the electron Fermi energy acquires its maximum value, namely when $\xi^{WS}=0$, the electron Fermi energy (\ref{tf_corr_fermi_1}) is the same as the one obtained from the uniform approximation which is independent of $N_p$. For smaller compressions, namely for $\xi^{WS}>0$ the electron Fermi energy deviates from the one given by the uniform approximation becoming $N_p$-dependent.

In Fig.~\ref{ANp1} we plot the Fermi energy of electrons, in units of the pion rest mass, as a function of the dimensionless parameter $\xi^{WS}$ and, as $\xi^{WS}\rightarrow 0$, the limiting value given by Eq.~(\ref{efa1}) is clearly displayed. 

In ref.~\cite{alcock}, in order to study the electrodynamical properties of strange stars, the ultra-relativistic Thomas-Fermi equation was numerically solved in  the case of bare strange stars as well as in the case of strange stars with a crust (see e.g. curves (a) and (b) in Fig.~6 of ref.~\cite{alcock}). In Fig.~6 of \cite{alcock} was plotted what they called the Coulomb potential energy, which we will denote as $V_{\rm Alcock}$. The potential $V_{\rm Alcock}$ was plotted for different values of the electron Fermi momentum at the edge of the crust. Actually, such potential $V_{\rm Alcock}$ is not the Coulomb potential $eV$ but it coincides with our function $e\hat V = eV+E_e^F$.
Namely, the potential $V_{\rm Alcock}$  corresponds to the Coulomb potential shifted by the the Fermi energy of the electrons. We then have from Eq.~(\ref{v01})
\begin{eqnarray}
e\hat V(\xi)=\left(\frac{9\pi}{4}\right)^{1/3}\frac{1}{\Delta} m_\pi c^2 \hat \phi(\xi)=V_{\rm Alcock}.
\label{efa3q}
\end{eqnarray}

This explains why in \cite{alcock}, for different values of the Fermi momentum at the crust the depth of the potential $V_{\rm Alcock}$ remains unchanged. Instead, the correct behaviour of the Coulomb potential is quite different and, indeed, its depth decreases with increasing of compression as can be seen in Fig.~\ref{efieldf}.
 
\section{Compressional energy of nuclear matter cores of stellar dimensions}\label{sec:6}

We turn now to the compressional energy of these  family of compressed nuclear matter cores of stellar dimensions each characterized by a different Fermi energy of the electrons. The kinematic energy-spectra of complete degenerate electrons, protons and neutrons are
\begin{eqnarray}
\epsilon^i(p) =  \sqrt{(pc)^2+m_i^2c^4} ,\quad p\le    P^F_i, \quad i=e,p,n. 
\label{local0}
\end{eqnarray}
So the compressional energy of the system is given by
\begin{align}
&{\mathcal E} = {\mathcal E}_{B}+{\mathcal E}_{e} + {\mathcal E}_{\rm em}\, ,\qquad
{\mathcal E}_{B} = {\mathcal E}_{p}+{\mathcal E}_{n}\, ,\\
&{\mathcal E}_{\rm i} = 2 \int_i\frac{d^3rd^3p}{(2\pi \hbar)^3} \epsilon^i(p)\, ,\quad i = e, p, n \, ,\qquad
{\mathcal E}_{\rm em} = \int  \frac{E^{2}}{8\pi}d^3r\, . \label{en2}
\end{align}
 
Using the analytic solution (\ref{2nd2s}) we calculate the energy difference between two systems, $I$ and $II$, 
\begin{equation}
\Delta{\mathcal E}={\mathcal E}(E^F_e(II))-{\mathcal E}(E^F_e(I)),
\label{loc51q}
\end{equation}
with $E^F_e(II) >
E^F_e(I) \geq 0$, at fixed $A$ and $R_c$. 

We first consider the infinitesimal variation of the total energy $\delta {\mathcal E}_{\rm tot}$ with respect to the infinitesimal variation of the electron Fermi energy $\delta E_e^F$
\begin{equation}
\delta{\mathcal E}=\left[\frac{\partial{\mathcal E} }{\partial N_p}\right]_{V^{WS}}\left[\frac{\partial N_p }{\partial E_e^F}\right] \delta E_e^F+\left[\frac{\partial{\mathcal E} }{\partial V^{WS} }\right]_{N_p}\left[\frac{\partial V^{WS} }{\partial E_e^F}\right] \delta E_e^F .
\label{loc1}
\end{equation}
For the first term of this relation we have
\begin{equation}
\left[\frac{\partial{\mathcal E} }{\partial N_p}\right]_{V^{WS}}=\left[\frac{\partial {\mathcal E}_{\rm p}}{\partial N_p}+\frac{\partial {\mathcal E}_{\rm n}}{\partial N_p}+\frac{\partial {\mathcal E}_{\rm e}}{\partial N_p}+\frac{\partial {\mathcal E}_{\rm em}}{\partial N_p}\right]_{V^{WS}}\simeq \left[E_p^F-E_n^F+E_e^F+\frac{\partial {\mathcal E}_{\rm em}}{\partial N_p} \right]_{V^{WS}} ,
\label{loc2}
\end{equation}
where the general definition of chemical potential $\partial \epsilon_i/\partial n_i = \partial {\cal E}_i/\partial N_i$ is used ($i=e$, $p$, $n$) neglecting the mass defect $m_n - m_p - m_e$. Further using the condition of the beta-equilibrium (\ref{npeq1e}) we  have
\begin{equation}
\left[\frac{\partial{\mathcal E}}{\partial N_p}\right]_{V^{WS}}=\left[\frac{\partial {\mathcal E}_{\rm em}}{\partial N_p} \right]_{V^{WS}}.
\label{loc3}
\end{equation}
For the second term of the Eq. (\ref{loc1}) we have
\begin{equation}
\left[\frac{\partial{\mathcal E} }{\partial V^{WS} }\right]_{N_p}=\left[\frac{\partial {\mathcal E}_{\rm p}}{\partial V^{WS}}+\frac{\partial {\mathcal E}_{\rm n}}{\partial V^{WS}}+\frac{\partial {\mathcal E}_{\rm e}}{\partial V^{WS}}+\frac{\partial {\mathcal E}_{\rm em}}{\partial V^{WS}}\right]_{N_p}=\left[\frac{\partial {\mathcal E}_{\rm e}}{\partial V^{WS}}\right]_{N_p}+\left[\frac{\partial {\mathcal E}_{\rm em}}{\partial V^{WS}}\right]_{N_p} ,
\label{loc4}
\end{equation}
since in  the process of increasing the electron Fermi energy namely, by decreasing the radius of the Wigner-Seitz cell, the system by definition maintains the same number of baryons $A$ and the same core radius $R_c$.

Now $\delta{\mathcal E}$ reads
\begin{equation}
\delta{\mathcal E}=\left\{\left[\frac{\partial {\mathcal E}_{\rm e}}{\partial V^{WS}}\right]_{N_p}\frac{\partial V^{WS} }{\partial E_e^F}+\left[\frac{\partial {\mathcal E}_{\rm em}}{\partial V^{WS}}\right]_{N_p}\frac{\partial V^{WS} }{\partial E_e^F}+\left[\frac{\partial {\mathcal E}_{\rm em}}{\partial N_p} \right]_{V^{WS}}\frac{\partial N_p }{\partial E_e^F}\right\} \delta E_e^F,
\label{loc5}
\end{equation}
so only the electromagnetic energy and the electron energy give non-null contributions.

From this equation it follows that
\begin{equation}
\Delta{\mathcal E}=\Delta{\mathcal E}_{\rm em}+\Delta{\mathcal E}_{\rm e}, 
\label{loc51}
\end{equation}
where $\Delta{\mathcal E}_{\rm em}={\mathcal E}_{\rm em}(E^F_e(II))-{\mathcal E}_{\rm em}(E^F_e(I))$ and $\Delta{\mathcal E}_{\rm e}={\mathcal E}_{\rm e}(E^F_e(II))-{\mathcal E}_{\rm e}(E^F_e(I))$.

In the particular case in which $E_e^F(II)=(E_e^F)_{max}$ and $E_e^F(I)=0$ we obtain
\begin{equation}
{\Delta\mathcal E} \simeq 0.75 \frac{3^{5/3}}{2}\left(\frac{\pi}{4}\right)^{1/3} \frac{1}{\Delta\sqrt{\alpha}}\left(\frac{\pi}{12}\right)^{1/6}N_p^{2/3}m_{\pi}c^2,
\label{local0q}
\end{equation}
which is positive. 

The compressional energy of a nuclear matter core of stellar dimensions increases with its electron Fermi energy as expected. 

\section{Conclusions}\label{sec:7}

We have generalized to the relativistic regime the classic work of Feynman, Metropolis and Teller, solving a compressed atom by the Thomas-Fermi equation in a Wigner-Seitz cell.

In the relativistic generalization the equation to be integrated is the relativistic Thomas-Fermi equation, also called the Vallarta-Rosen equation \cite{vallarta}. The integration of this equation does not admit any regular solution for a point-like nucleus and both the nuclear radius and the nuclear composition have necessarily  to be taken into account \cite{ruffini, ruffinistella81}. This introduces a fundamental difference from the non-relativistic Thomas-Fermi model where a point-like nucleus was adopted.

As in  previous works \cite{ruffini, ruffinistella81, rrx200613}  the protons in the nuclei have been assumed to be at constant density, the electron distribution has been derived by the Thomas-Fermi relativistic equation and  the neutron component has been derived by   the  beta equilibrium between neutrons, protons and electrons.

We have also examined for completeness the relativistic generalization of the Thomas-Fermi-Dirac equation by taking into due account the exchange terms \cite{dirac30}, adopting  the general approach of Migdal, Popov and Voskresenskii \cite{migdal77}, and shown that these effects, generally small, can be neglected in the relativistic treatment.

There are marked differences between the relativistic and the non-relativistic treatments. 

The first and the most general one is that the existence of a finite size nucleus introduces necessarily a limit to the compressibility: the dimension of the Wigner-Seitz cell can never be smaller then the nuclear size. Consequently the electron Fermi energy which in the non-relativistic approach can reach arbitrarily large values, reaches in the present case a perfectly finite value whose expression has been given in analytic form. There are in the literature many papers adopting a relativistic treatment for the electrons together with a point-like approximation for the nucleus, which is clearly inconsistent (see e.g. \cite{gilles1, gilles2}). 

The second is the clear difference of the electron distribution as a function of the radius and of the nuclear composition as contrasted to the uniform approximation often adopted in the literature (see e.g.\cite{mishustin}) which we have explicitly shown in the Fig.~\ref{fmtw} of Sec. \ref{sec:4}. Inferences based on the uniform approximation are not appropriate both in the relativistic and in the non-relativistic regime. 

The third, one of the most relevant, is that the relativistic Feynman-Metropolis-Teller treatment allows to treat globally and in generality the electrodynamical interaction within the atom and the relativistic corrections leading to a softening of the dependence of the electron Fermi energy on the compression factor, as well as a gradual decrease of the exchange terms in proceeding from the non-relativistic to the fully relativistic regimes. It is then possible to derive, as shown in Table \ref{xx1} of Sec.~\ref{sec:4} a consistent equation of state for compressed matter which overcomes some of the difficulties of existing treatments describing the electrodynamical effect by a sequence of approximations which have lead to the occurrence of unphysical regimes e.g. the existence of negative pressure as in the Salpeter approach. As a direct application of this treatment we have reconsidered the study of white dwarfs within the relativistic Feynman, Metropolis, Teller approach and evaluate their effects on the value of the radii, of the masses of the equilibrium configurations as well as on the numerical value of the critical mass \cite{inpreparation}. We have there compared and contrasted the results obtained by Chandrasekhar with a uniform approximation with the ones obtained by the equation of state of Salpeter and the ones following from the treatment presented in this article.

We have then extrapolated these results to the case of nuclear matter cores of stellar dimensions for $A\approx(m_{\rm Planck}/m_n)^3 \sim 10^{57}$ or $M_{core}\sim M_{\odot}$. The aim here is to explore the possibility of obtaining for these systems a self-consistent solution presenting global and not local charge neutrality. The results generalize the considerations presented in the previous article   corresponding to a nuclear matter core of stellar dimensions with null Fermi energy of the electrons \cite{prl}. The ultra-relativistic approximation allows to obtain analytic expressions for the fields. The exchange terms can in this approximation be safely neglected. An entire family of configurations exist with values of the Fermi energy of the electrons ranging from zero to a maximum value $(E_e^F)_{max}$ which is reached when the Wigner- Seitz cell coincides with the core radius. The configuration with $E_e^F=(E_e^F)_{max}$ corresponds to the configuration with $N_p=N_e$ and $n_p=n_e$. For this limiting value of the Fermi energy  the system fulfills both the global and the local charge neutrality and correspondingly no electrodynamical structure is present in the core. 
All the other configurations presents overcritical  electric fields close to their surface. The configuration with $E_e^F=0$ has the maximum value of the electric field at the core surface, well above the critical value $E_c$ (see Fig. \ref{efieldf}, Fig.~\ref{efieldf1} and Fig.~\ref{ANp} of Section \ref{sec:5}). All these cores with overcritical electric fields are  stable against the vacuum polarization process due to the Pauli blocking by the degenerate electrons \cite{physrep}. We have also compared and contrasted our treatment of the relativistic Thomas-Fermi solutions to the corresponding one addressed in the framework of strange stars \cite{alcock} pointing out in these treatments some inconsistency in the definition of the Coulomb potential.

We have finally compared the compressional energy of configurations with selected values of the electron Fermi energy. In both systems of the compressed atoms and of the nuclear matter cores of stellar dimensions a maximum value of the Fermi energy has been reached corresponding to the case of Wigner-Seitz cell radius $R_{WS}$ coincident with the core radius $R_c$.

Both  problems  considered, the one of a compressed atom and the one of compressed nuclear density core of stellar dimensions, have been treated by the solution of the relativistic Thomas-Fermi equation and by enforcing the condition of beta equilibrium. They are theoretically well defined and, in our opinion, a necessary step in order to approach the more complex problem of a neutron star core and its interface with the neutron star crust. 

Neutron stars are composed of two sharply different components: the liquid core at nuclear and/or supra-nuclear density consisting of neutrons, protons and electrons and a crust of degenerate electrons in a lattice of nuclei \cite{baym, harrisonweel} and possibly of free neutrons due to neutron drip when this process occurs (see e.g. \cite{baym}). Consequently, the boundary conditions for the electrons at the surface of the neutron star core will have generally a positive value of the electron Fermi energy in order to take into account the compressional effects of the neutron star crust on the core \cite{letter3}. The case of zero electron Fermi energy corresponds to the limiting case of absence of the crust.

In a set of interesting papers \cite{glendenning92, glendenning95, glendenning97, glendenning99, glendenning00, glendenning01} Glendenning has relaxed the local charge neutrality condition for the description of the mixed phases in hybrid stars. In such configurations the global charge neutrality condition, as opposed to the local one, is applied to the limited regions where mixed phases occur while in the pure phases the local charge neutrality condition still holds. We here generalize Glendenning's considerations by looking to a violation of the local charge neutrality condition on the entire configuration, still keeping its overall charge neutrality. This effect does not occur in processes occurring just locally, but it requires the global description of the equilibrium configuration. As exemplified in \cite{ inpreparationa, inpreparationb}, where gravitational effects are taken into account both in the $T=0$ and $T\neq0$ cases, the local properties need the previous knowledge of the entire equilibrium configuration and therefore the description is necessarily global.

In conclusion the analysis of compressed atoms following the relativistic Feynman, Metropolis, Teller treatment presented in the first part of this article has important consequences in the determination of the mass-radius relation of white dwarfs \cite{inpreparation} leading to the possibility of a direct confrontation of these results with observations, in view of the current great interest for the cosmological implications of the type Ia supernovae \cite{phillips,riess98,perlmutter, riess04}. The results presented in the second part of this article on nuclear matter cores of stellar dimensions evidence the possibility of having the existence of critical electromagnetic fields in the interface of the core and the neutron star crust. The results here obtained in a simplified but rigorous approach of the application of the relativistic Feynman, Metropolis, Teller treatment to the constant density cores in beta equilibrium  conform  to the prediction of Baym, Bethe and Pethick \cite{baym}. This treatment has been further extended to the case in which a self-gravitating system of degenerate neutrons, protons and electrons is considered within the framework of relativistic quantum statistics and Einstein-Maxwell equations \cite{letter3} and to the case in which also strong interactions are present \cite{strong}.  

\begin{acknowledgments}

We thank the anonymous referee for his/her helpful comments and suggestions.

\end{acknowledgments}

\end{document}